\documentclass[trackchanges,twocolumn]{aastex701}

\usepackage{fontawesome}
\usepackage{hyperref}

\newcommand{\githubicon}{{\color{black}\faGithub}}

\newcommand{\prot}{$P_\textrm{rot}$~}

\begin{document}

\title{KRONOS I: The $1{-}2.8\mu$m JWST Transmission Spectrum of the 23 Myr V1298~Tau~c}
\shorttitle{The Young Super-Earth Progenitor V1298~Tau~c}
\shortauthors{Murphy et al. 2026}
\received{April 13, 2026}
\revised{June 2, 2026}
\accepted{June 3, 2026}

\author[orcid=0000-0002-8517-8857]{Matthew M. Murphy}
\affiliation{Department of Physics and Astronomy, Michigan State University, East Lansing, MI, USA}
\email[show]{mmmurphy@msu.edu}  

\author[orcid=0000-0001-8236-5553]{Matthew C. Nixon}
\altaffiliation{51~Pegasi~b Fellow}
\affiliation{School of Earth and Space Exploration, Arizona State University, Tempe, AZ, USA}
\email[]{matthewnixon@asu.edu}

\author[orcid=0000-0002-9464-8101]{Adina D. Feinstein}
\affiliation{Department of Physics and Astronomy, Michigan State University, East Lansing, MI, USA}
\email[]{adina@msu.edu}

\author[orcid=0000-0003-0156-4564]{Luis Welbanks}
\affiliation{School of Earth and Space Exploration, Arizona State University, Tempe, AZ, USA}
\email[]{luis.welbanks@asu.edu}

\author[orcid=0000-0002-7119-2543]{Girish M. Duvvuri}
\affiliation{Department of Physics and Astronomy, Vanderbilt University, Nashville, TN 37235, USA}
\email[]{girish.duvvuri@gmail.com}

\author[orcid=0009-0000-6113-0157]{Saugata Barat}
\affiliation{Kavli Institute of Astronomy, Massachusetts Institute of Technology, Cambridge, USA}
\email[]{saugata@mit.edu}

\author[orcid=0000-0002-3627-1676]{Benjamin V. Rackham}
\affiliation{Kavli Institute for Astrophysics and Space Research, Massachusetts Institute of Technology, Cambridge, MA 02139, USA}
\affiliation{Department of Earth, Atmospheric and Planetary Science, Massachusetts Institute of Technology, 77 Massachusetts Avenue, Cambridge, MA 02139, USA}
\email[]{brackham@mit.edu}

\author[orcid=0000-0002-0726-6480]{Darryl Z. Seligman}
\affiliation{Department of Physics and Astronomy, Michigan State University, East Lansing, MI, USA}
\email[]{dzs@msu.edu}

\author[orcid=0000-0002-3328-1203]{Michael Radica}
\altaffiliation{NSERC Postdoctoral Fellow}
\affiliation{Department of Astronomy \& Astrophysics, University of Chicago, 5640 South Ellis Avenue, Chicago, IL 60637, USA}
\email[]{radicamc@uchicago.edu}

\author[]{Ian J. M. Crossfield}
\affiliation{Department of Physics and Astronomy, University of Kansas, Lawrence, KS, USA}
\email[]{ianc@ku.edu}

\author[orcid=0000-0002-1002-3674]{Kevin France}
\affiliation{Laboratory for Atmospheric and Space Physics, University of Colorado Boulder, Boulder, CO 80309}
\email[]{kevin.france@colorado.edu}

\author[orcid=0000-0002-4881-3620]{John H. Livingston}
\affiliation{Astrobiology Center, 2-21-1 Osawa, Mitaka, Tokyo 181-8588, Japan}
\affiliation{National Astronomical Observatory of Japan, 2-21-1 Osawa, Mitaka, Tokyo 181-8588, Japan}
\affiliation{Astronomical Science Program, The Graduate University for Advanced Studies (SOKENDAI), 2-21-1 Osawa, Mitaka, Tokyo 181-8588, Japan}
\email[]{john.livingston@nao.ac.jp}

\author[orcid=0000-0003-2279-4131]{Jonathan Lunine}
\affiliation{Jet Propulsion Laboratory, California Institute of Technology}
\affiliation{Division of Geologic and Planetary Sciences, Caltech}
\email[]{jlunine@caltech.edu}

\author[0000-0002-4671-2957]{Rafael Luque}
\affiliation{Instituto de Astrof\'isica de Andaluc\'ia (IAA-CSIC), Glorieta de la Astronom\'ia s/n, 18008 Granada, Spain}
\email[]{rluque@iaa.es}

\author[orcid=0000-0001-8504-5862]{Catriona Murray}
\affiliation{Department of Astrophysical and Planetary Sciences, University of Colorado Boulder, Boulder, CO 80309, USA}
\email[]{catriona.murray@colorado.edu}

\author[orcid=0000-0003-1622-1302]{Sagnick Mukherjee}
\altaffiliation{51~Pegasi~b Fellow}
\affiliation{School of Earth and Space Exploration, Arizona State University, Tempe, AZ, USA}
\email[]{smukhe50@asu.edu}

\author[orcid=0009-0007-8422-1589]{Biruk Nardos}
\affiliation{School of Earth and Space Exploration, Arizona State University, Tempe, AZ, USA}
\email[]{bnabebe@asu.edu}

\author[orcid=0009-0008-8016-6591]{Sydney Petz}
\affiliation{Department of Physics and Astronomy, Michigan State University, East Lansing, MI, USA}
\email[]{petzsydn@msu.edu}

\author[0000-0001-9289-0570]{Hinna Shivkumar}
\affiliation{Anton Pannekoek Institute for Astronomy, University of Amsterdam, Science Park 904, 1098 XH, Amsterdam, The Netherlands}
\email[]{h.shivkumar@uva.nl}

\begin{abstract}
While recent JWST observations of mature super-Earths and sub-Neptunes have frequently revealed featureless transmission spectra, their inflated progenitors offer a unique window into understanding their primordial compositions. As part of the Keys to Revealing the Origin and Nature Of sub-neptune Systems (KRONOS) JWST program, we present the NIRISS/SOSS transmission spectrum of V1298~Tau~c, a $\sim$23~Myr super-Earth progenitor orbiting a young Solar analog. We detect H$_2$O in V1298~Tau~c's atmosphere with a $\log_{10}$ volume mixing ratio of $-1.83^{+0.68}_{-0.77}$, but no additional molecules from these data alone. We find consistent results for the planetary atmospheric properties in both retrievals with and without informed priors on stellar heterogeneities based on the observed stellar spectrum. We infer an atmospheric metallicity [O/H] of $14.8^{+56.0}_{-12.28}\times$ the solar value. This metallicity is similar to literature measurements for other young planets, including its massive outer companion V1298~Tau~b. In contrast, this measured metallicity is systematically lower than the metallicities of mature planets of similar mass and temperature. Altogether, these results provide tentative but growing evidence that the exoplanet mass--metallicity relation evolves with planetary age. 
\end{abstract}
% note 250 allowed words.

\section{Introduction} \label{sec:intro}

Super-Earth- and sub-Neptune-sized exoplanets ($\sim$1.2--5~$R_\oplus$) appear to be the most typical output of planet formation at short-period orbits \citep{howard2012, howard2025, fressin2013, fulton2017, bean2021}. Despite their prevalence and community attention in the JWST era, little is still known about their nature and diversity. Several transmission spectra of sub-Neptunes ($R_p > 1.7 R_\oplus$) have evidence of H$_2$O, CH$_4$, CO$_2$, and SO$_2$ \citep{mahdu2023, benneke2024_toi270d, beatty2024_gj3470, piaulet2024_GJ9827d, schlawin2024_gj1214, davenport2025, roy2026_lp79118}, while others have remained featureless due to the presence of aerosols, contamination from stellar heterogeneities, and/or the lack of an extended envelope \citep[e.g.,][]{cadieux2024_lhs1140, wachiraphan2025_LTT1445Ab, redai2025_gj357b, ahrer2025_gj3090}. The smaller super-Earths ($R_p < 1.7 R_\oplus$) have been significantly more challenging to characterize, often due to featureless transmission spectra \citep[e.g.][]{may2023_gj1132, taylor2025_gj357b, Moran2023_GJ486b, scarsdale2024_L9859c, luque2025_TOI1685b, alam2025_l1689}, leading to ambiguity about their atmospheric properties.

Nearly all of the previously observed super-Earths and sub-Neptunes are mature ($>$1~Gyr) and therefore represent the near-end state of significant evolutionary processing. It is expected that this population was heavily sculpted by atmospheric evolutionary processes, such as photoevaporation \citep{Lammer2003, Yelle2004, Baraffe2006, MurrayClay2009, Jackson2012, Owen2013, Lopez2013, Jin2014, Chen2016, Owen2017}, core-powered mass loss \citep{Ginzburg2016, Gupta2019, Gupta2020, Misener2021, Owen2024, Misener2025}, planetesimal accretion \citep{Seligman2022,ODonovan2026}, atmosphere-interior interactions \citep{schlichting22, nixon25}, and giant impacts \citep{Inamdar2016}, earlier in their lifetimes. Therefore, we may only be observing the diverse remnants of a modified population of planets. Atmospheric characterization of young ($\lesssim$100~Myr) transiting exoplanets thus provides a unique opportunity to capture this evolution, and understand the primordial nature of super-Earths and sub-Neptunes.  

The Keys to Revealing the Origin and Nature Of sub-neptune Systems (KRONOS) program (JWST GO 5959; co-PIs: Feinstein \& Welbanks) observed three young multiplanet systems hosting sub-Neptune and super-Earth progenitors: V1298~Tau \citep[$\sim$23~Myr;][]{david2019_v1298tau, david19_b}, TOI-451 \citep[$\sim$100~Myr;][]{newton21}, and TOI-2076 \citep[$\sim$200~Myr;][]{hedges21, hedges22, osborn22, barber25}. In a forthcoming series of articles, we will present and compare the $\sim$1--5~$\mu$m JWST NIRISS/SOSS and NIRSpec/G395H transmission spectra of two or more planets in each system. This work focuses on the first subset of the V1298~Tau observations.

V1298 Tau is a young, Sun-like ($\sim$1.1\,M$_\odot$, 1.3\,R$_\odot$) K0 dwarf hosting four confirmed transiting exoplanets \citep{david2019_v1298tau, david19_b, feinstein2022_v1298tauTESS, livingston2026_v1298tau}. Based on astrometric data \citep{oh2017} and strong X-ray emission from the \textit{ROSAT} All-Sky Survey \citep{wichmann96}, V1298~Tau is likely a member of a spatially distributed subgroup of the Taurus--Auriga Association known as Group 29. Further analyses using gyrochronology (\prot=~2.9~days), show evidence of excess UV emission from the \textit{GALEX} mission \citep{findeisen2010}, and isochronal fitting have constrained the age of the system to $20-30$~Myr \citep{david19_b, suarez_mascareno_2022, johnson2022}; we adopt the age of $t_\textrm{age} = 23 \pm 4$~Myr from \cite{david19_b}. This young age suggests the V1298~Tau planets should still have a significant portion of their primordial atmospheres, but are experiencing intense atmospheric evolution \citep{Owen2017}. 

V1298~Tau~c is the innermost known transiting planet in the V1298~Tau system, with an orbital period of approximately 8.25~days \citep{feinstein2022_v1298tauTESS, livingston2026_v1298tau}. Recent long-baseline transit timing variation (TTV) measurements by \cite{livingston2026_v1298tau} constrained V1298~Tau~c's mass and radius to 4.7 $\pm$ 0.6~M$_\oplus$ and 5.08 $\pm$ 0.37~R$_\oplus$, respectively. This combination yields one of the lowest bulk densities (0.20 $\pm$ 0.05~g/cm$^3$) of all known transiting exoplanets, but also the highest bulk density among the other V1298~Tau planets. These properties all together suggest V1298~Tau~c is likely experiencing the most fervent evolution in the system. In fact, photoevaporation models predict V1298~Tau~c will eventually attain a final radius $<$2.5\,$R_\oplus$ \citep{poppenhaeger21, livingston2026_v1298tau}. Thus, V1298~Tau~c offers a prime case study of the precursor to our Galaxy's common super-Earth and sub-Neptune population.

Here we present and interpret the $\sim$$1{-}2.8\,\mu$m JWST NIRISS/SOSS transmission spectrum of V1298 Tau c. We observed at least six starspot crossings during the transit, which we leverage to explore different methods for modeling them during light curve fitting, and infer the photospheric and starspot properties of V1298~Tau. In this work, we focus primarily on the atmospheric properties of V1298~Tau~c. We will analyze the starspot properties from these data in detail in a follow-up work \citep{murphy2026_spots}. The remainder of the manuscript is organized as follows. In Section~\ref{sec:obs}, we describe our observations and data reduction. In Section~\ref{sec:methods}, we describe our light curve fitting techniques. We present the broadband light curve fits in Section~\ref{sec:bbfits}, the transmission spectrum of V1298 Tau c in Section~\ref{sec:spectrum}, and the stellar spectrum of V1298 Tau in Section~\ref{sec:stellarspectrum}. We interpret our observations through atmospheric modeling, and present these results in Section~\ref{sec:atmomodeling}. Then, we discuss and contextualize the evolutionary history of V1298 Tau c in Section~\ref{sec:formation}. Finally, we summarize our work in Section~\ref{sec:summary}.

This manuscript aims to improve reproducibility and transparency of scientific research. All of the data presented are hosted on GitHub. Every figure caption includes a GitHub icon (\githubicon), which links to an interactive Jupyter notebook used to create that figure.

%%% ===================================================================== %%%
\section{Observations and Data Reduction} \label{sec:obs}
\subsection{Observation Details} \label{subsec:methods_obsdetails}
As part of the KRONOS program JWST GO 5959, we observed one transit of the exoplanet V1298~Tau~c using JWST's Near Infrared Imager and Slitless Spectrograph \citep[NIRISS; ][]{niriss1_overview} on 2025 September 6 starting at UTC 02:05:25. We acquired directly on the target using the \texttt{SOSSFAINT} acquisition mode with the F480M filter and \texttt{NISRAPID} readout pattern, taking a single integration of three groups. Then, we switched to the Single Object Slitless Spectroscopy mode \citep[SOSS; ][]{niriss3_soss} with the \texttt{SUBSTRIP96} subarray and \texttt{NISRAPID} readout pattern to perform time series spectroscopy from $\sim$0.85--2.83\,$\mu$m continuously. Our subarray choice was driven by V1298~Tau's relatively bright magnitude ($V$=10.115, $J$=8.687, $K$=8.094) and, as a result, we only have access to SOSS's first spectral order. We took 2598 integrations, each consisting of five groups, for a total exposure/visit duration of $\sim$9.6\,hr. This number of groups was sufficient to avoid saturation across the entire detector. The transit duration was 4.78~hr (Table~\ref{tab:bbfit_params}), and we had 1.92~hr and 2.87~hr of pre- and post-transit baseline, respectively.

\subsection{Data Reduction} \label{subsec:methods_datareduction}

We reduced the NIRISS time series data using version 2.3.1 of the open-source \texttt{exoTEDRF} pipeline \citep{feinstein_early_2023, radica2023_wasp96, exotedrf}. The detailed data reduction methods are well described elsewhere \citep[e.g.,][]{radica2024_ltt9779, radica_promise_20225, radica_super-solar_2026}, and here we summarize the key details relevant to our dataset. 

Several steps of the pipeline can optionally use  outputs from later steps as informed inputs. These outputs include (i) pixel masks, (ii) estimates of which integrations are out of transit, and (iii) estimates of the normalized spectroscopic light curves. We therefore performed an initial reduction to generate these informed data, then saved the relevant outputs and repeated the reduction. We identified integrations 0--520 and 1800--2598 as being out of transit. We followed the standard Stage 1 steps, inputting the hot pixel map of the initial run to the \texttt{DQInit} step. We performed background subtraction here at the group level, uniformly rescaling the STScI \texttt{SUBSTRIP96} background model\footnote{\url{https://github.com/radicamc/exoTEDRF/blob/main/files/model_background96.npy}} to match the flux level of the observations. We passed bad pixel masks and a time series estimate from the initial run as well as the STScI background image into the \texttt{OneOverF} step. We ran this step using the \texttt{scale-achromatic} method with an inner mask width of 40 pixels. Then, we run the \texttt{LinearityStep} to correct for detector nonlinearity, which successfully reduced the group-to-group differences from a maximum of $\sim$12\,data numbers (DN) relative to the mean to $<$0.7\,DN.

Next, we followed the standard Stage 2 routine, including repeating background subtraction at the integration level. Near the end of Stage 2, we perform a principal component analysis (PCA) on the series of 2D images to identify and remove known systematic artifacts, as demonstrated and validated in \citet{radica_super-solar_2026}. The only component we removed was the beating pattern associated with the telescope's thermal control system \citep{coulombe2023_wasp18}. Through the PCA, we also identified a possible mirror-tilt event \citep[e.g.,][]{alderson_early_2023, radica2023_wasp96} around integration 2390, approximately 2.2 hours after transit egress and only 45 minutes before the end of the observation. Rather than attempt to correct for this via PCA reconstruction, we simply clip all integrations after this jump event from each light curve fit. 

\begin{figure}
    \centering
    \includegraphics[width=1\linewidth]{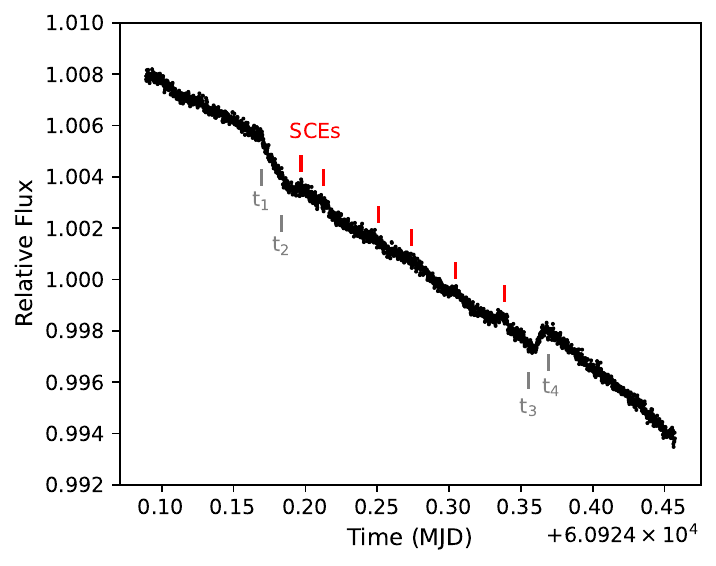}
    \caption{The band-integrated ($\sim$0.85--2.83\,$\mu$m) transit light curve of V1298~Tau~c, normalized to the visit median flux, as observed using JWST/NIRISS SOSS on UT 2025 September 6. The transit contact points (t$_1$--t$_4$) are labeled. We see a significantly curved out-of-transit baseline and numerous starspot crossing events (SCEs). 
    \href{https://github.com/kronos-jwst/KRONOS-I-V1298-Tau-c-NIRISS/blob/main/Create_Figure1.ipynb}{\githubicon}
    }
    \label{fig:raw_bblc}
\end{figure}

In Stage 3, we performed a simple box extraction with a half-width of 20 pixels around the identified trace, which we ensured captured $\sim$99\% of the signal per column. We  flux calibrated the 1D spectra in parallel for use in inferring the stellar properties of V1298~Tau, using the \texttt{flux\_calibrate\_soss} step. We present the extracted, band-integrated light curve in Figure~\ref{fig:raw_bblc}.

Performing multiple independent reductions on the same dataset has been the precedent for analyses of JWST transit observations. This practice was useful for identifying features in the observed time series as robustly astrophysical or imprinted by choices made in the reduction process early in the mission's lifetime. Due to these efforts, best practices for the reduction process are now relatively well known, and the results from independent pipelines generally agree to high fidelity \citep[e.g.,][]{cartermay2024, beatty2024_gj3470, radica2024_ltt9779, schlawin2024_gj1214, schmidt2025_k218b, luque2025_k218b}. While there are still minor differences between reduction pipelines, our light curves are dominated by starspot crossing events (SCEs) whose effects should dwarf reduction-level choices. 
For example, transmission spectra derived from independent reductions of the same data but similar light curve fitting methods typically agree within the transit depth uncertainty \citep[e.g.,][]{feinstein_early_2023, davenport2025, ahrer2025_gj3090, radica_super-solar_2026}, whereas differences due to choices on SCEs or limb darkening can be several times larger \citep[e.g.,][]{murray2025_SCEeffects, mercier2025_limbdarkening}. As such, the remainder of the analysis is focused on testing light-curve fitting techniques, further discussed in Section~\ref{sec:methods}.
Therefore, in this and future KRONOS program papers, we choose to streamline the Stage 1--3 reduction process by using only one pipeline (\texttt{exoTEDRF}), and instead conduct in-depth intercomparisons in transmission spectra that have been fit with different light-curve fitting routines. 

%%% ===================================================================== %%%
\section{Light Curve Fitting} \label{sec:methods}

Young stars like V1298~Tau are statistically more active compared to their older main-sequence counterparts \citep{Grankin1999_youngspottystar, GullySantiago2017_youngspottystar, Feinstein2020_youngstarflares, Namekata2025_youngstarCMEs}. Previous observations of the V1298~Tau system have captured numerous starspot crossings \citep{feinstein2022_v1298tauTESS}, flares \citep{barat2025_v1298taub}, short-term variability in H$\alpha$ absorption \citep{feinstein2021_v1298tauGemini}, and long-term stellar variability \citep{david2019_v1298tau, feinstein2022_v1298tauTESS, suarez_mascareno_2022}. Observations of young exoplanets must, therefore, carefully consider the impacts of stellar activity, particularly stellar surface heterogeneities like starspots. At the same time, these observations provide a unique opportunity to study the properties of both occulted and unocculted heterogeneities on young stars. 

Based on the broadband light curve, the out-of-transit baseline is significantly nonlinear (Figure~\ref{fig:raw_bblc}). We also see numerous starspot crossings events\footnote{These may actually be clusters of starspots or generally active regions but, for brevity, we refer to them as individual starspots.} (SCEs) during the transit. These motivate a comprehensive light-curve analysis to understand what impact these SCEs have on the observed transmission spectrum, and enable inference of the properties of the starspots themselves. To this end, we designed four methods to account for the SCEs in the light-curve fitting: 1) masking the SCEs, 2) modeling the SCEs as Gaussian profiles, 3) modeling the SCEs using the \texttt{fleck} starspot model, and 4) modeling the SCEs using a Gaussian process model. Section~\ref{subsec:lcfitting_method} describes our overall methodology common to all fits, then the individual details of these methods are described in Sections~\ref{subsec:lcfitting_case1mask}--\ref{subsec:lcfitting_case4gp}. We perform these tests on both the broadband and spectroscopic light curves, which we create at resolutions of $R$=50, 100, and 300. In this work, we adopt a bin-then-fit approach by fitting prebinned spectroscopic light curves, rather than binning down the transmission spectrum derived from native resolution spectroscopic light curves \citep{cartermay2024}.

\subsection{Common Modeling and Sampling Method} \label{subsec:lcfitting_method}

We used a custom \texttt{python} routine to model each transit light curve as the product of a transit model $T\left(\theta, t\right)$, an SCE model $F_{\mathrm{SCE}}\left(\theta, t\right)$, and an out-of-transit baseline model $S\left(\theta, t\right)$, which encapsulates both instrumental and stellar long-term trends. All three models are functions of the parameter array $\theta$, which includes astrophysical and baseline parameters, and time $t$. 

To model the out-of-transit baseline, we used a cubic polynomial of the form
\begin{equation}
    S\left(t\right) = a_3 ~ x^3 + a_2 ~ x^2 + a_1 ~ x + a_0, \label{eqn:polynomial}
\end{equation}
where $x \equiv t - t_{\text{median}}$ ($t_{\text{median}}$ is the visit median time) and $a_i$ are coefficients for the $i$th order. We tested other polynomials of increasing complexity from two to five terms, but the cubic model was preferred via the Bayesian Information Criterion (BIC) (see Table~\ref{tab:bics}). 

We first fit the broadband light curves to refine the orbital parameters of V1298~Tau~c and most of the SCE parameters, which would later be fixed when fitting the spectroscopic light curves. For all tests, we fit for the local time of conjunction $t_c$, the semi-major axis $a$ parameterized as $\log_{10}\left(a/R_\star\right)$, the orbital inclination $i$ as $\cos\left(i\right)$, the planet-star radius ratio $R_{\mathrm{p}} / R_\star$, the limb darkening coefficients $u_1$ and $u_2$ (see Sec.~\ref{subsec:lcfitting_limbdarkening} for details), the baseline polynomial terms $a_3$--$a_0$, and a multiplicative relative flux uncertainty scaling term $\sigma_{\mathrm{scaler}}$. These parameters are listed in Table~\ref{tab:bbfit_params}. The latter term ensures the uncertainty on each integration is comparable to the scatter of the light curve and data--model residuals. Then, we include additional parameters for the SCE models described further in Sections~\ref{subsec:lcfitting_case1mask}--\ref{subsec:lcfitting_case4gp}. We fix the orbital period to 8.2438\,days \citep{feinstein2022_v1298tauTESS}. We set Gaussian priors on $a/R_\star$ and $i$ based on \cite{david2019_v1298tau}. For all other parameters, we use  uninformed uniform priors, which are unbounded except for limiting $R_{\mathrm{p}} / R_\star$ as well as the limb darkening coefficients to their physically plausible ranges of (0, 1) inclusively. We also enforce that $\sigma_{\mathrm{scaler}} > 0$.

We fit the broadband light curves with Markov Chain Monte Carlo (MCMC) sampling using \texttt{emcee} \citep{emcee}, running the sampling for a 10,000 step burn-in period followed by a 50,000 step sampling period. We set the number of walkers to three times the number of free parameters in each case. We verified convergence based on the auto-correlation times using a minimum threshold of 30$\times$, and inspection of the walker time series and posterior distributions. 

When fitting the spectroscopic light curves, we fixed each of the orbital parameters except $t_\mathrm{c}$ to the broadband-derived values from that respective case. We leave $t_c$ free as its chromatic variation can provide evidence for evening-morning limb asymmetry, which we discuss further in Section~\ref{subsec:atmomodeling_discussion_LAdiscussion}. Then, we fit for $t_{\mathrm{c}}$, $R_{\mathrm{p}} / R_\star$, $u_1$, $u_2$, $a_3$--$a_0$, $\sigma_{\mathrm{scaler}}$, and the wavelength-dependent SCE properties mentioned in Sections~\ref{subsec:lcfitting_case2gaussian}--\ref{subsec:lcfitting_case4gp}. We sampled each spectroscopic channel independently, running each for a 1,000 step burn-in period followed by a 2,500 step sampling period. 

\subsection{Test Case 1: Masking Spot Crossing Events} \label{subsec:lcfitting_case1mask}

As the first and most conservative approach, we masked the apparent SCEs in each light curve. The full light curve model $F_{\mathrm{obs}}$ is therefore simply
\begin{equation}
    F_{\mathrm{obs, 1}} \left(t\right) = T\left(\theta, t\right) \times S\left(\theta, t\right) \,.\label{eqn:case1model}
\end{equation}
No new parameters beyond those listed in Section~\ref{subsec:lcfitting_method} are required. Here, we used \texttt{batman} \citep{batman} for the transit model $T\left(\theta, t\right)$. 

We identified the affected regions by eye as relative flux values that deviate from the expected smooth, lower floor of the transit, and were agnostic to whether such deviations could be from one or more individual starspot crossings. We identified and masked integrations 658--850, 970--1300, 1370--1450, and 1550--1700 in both the broadband and spectroscopic light curve fits. 

\subsection{Test Case 2: Modeling Spot Crossing Events as Gaussian Profiles} \label{subsec:lcfitting_case2gaussian}

As a more flexible but physically agnostic approach, we modeled the SCEs as Gaussian profiles. The relative flux contribution from SCEs is 
\begin{equation}
    F_{{\mathrm{SCE}, 2}} = 1 + \sum_{i} A_i \,\, \textrm{exp}\bigg({-\frac{\big(t - b_i\big)}{2 \sigma_i^2}}\bigg)\,, \label{eqn:GaussianFunc}
\end{equation}
where $A_i$ is the relative flux amplitude, $b_i$ sets the center of the Gaussian in units of time (i.e., the mid-point of the SCE), and $\sigma_i$ sets the width of the Gaussian in units of time (i.e., the duration of the SCE) for the $i$th SCE. This contribution is multiplicative with the spot-free transit model, so the full light curve model is
\begin{equation}
    F_{\mathrm{obs, 2}} \left(t\right) = T\left(\theta, t\right) \times S\left(\theta, t\right) \times F_{\mathrm{SCE}, 2} \,.\label{eqn:case2model}
\end{equation}
We again used \texttt{batman} for the transit model. 

We prescribed each SCE a separate $A$, $b$, and $\sigma$ value, and included these as additional free parameters in the broadband light curve fit. Since we expect only the amplitude---linked to the starspot contrast---to be wavelength-dependent, we then fixed the values of $b$ and $\sigma$ in the spectroscopic fits, and fit only $A$ as a function of wavelength. We use  uninformed uniform priors on each of these SCE parameters, and allow $A$ to be negative to allow for facular crossings in addition to spots.

\subsection{Test Case 3: Modeling Spot Crossing Events using \texttt{fleck}} \label{subsec:lcfitting_case3fleck}

As a more physical extension of Case~2, we adopted the \texttt{fleck} package \citep{fleck} to model the SCEs. For our purposes, \texttt{fleck} semi-analytically calculates the change in the transit light curve due to a planet occulting a starspot of given contrast $c$, radius $r_{\mathrm{spot}}$, and position (latitude $lat$ and longitude $lon$) on the stellar surface. \texttt{fleck} implements its own transit model, $T_{\mathrm{fleck}}$ based on the framework of \texttt{batman}, which automatically includes the SCE profiles. The full light curve model is simply 
\begin{equation}
    F_{\mathrm{obs, 3}} \left(t\right) = T_{\mathrm{fleck}} \left(\theta, t\right) \times S\left(\theta, t\right) \,.\label{eqn:case3model}
\end{equation}

\texttt{fleck} enforces the same contrast value between each starspot, but we still allow each SCE to have its own radius and position, which were included as additional free parameters in the broadband fit. We then fixed the radius and position of each spot in the spectroscopic fits, leaving only the contrast to be fit as a function of wavelength. We use uninformed uniform priors on each of these parameters. However, we enforce that the radii are non-negative and limit the spot longitudes to (-90$^\circ$, 90$^\circ$) to keep them on the visible disk.

We did not model the effects of stellar rotation using \texttt{fleck}. \texttt{fleck} predicts the out-of-transit baseline based on the input starspots and stellar rotation period. The predicted baseline is primarily driven by unocculted spots which rotate into or out of view during the observation. However, our data insufficiently constrains the number, size, or position of unocculted spots. Combined with V1298~Tau's short $\sim$2.9~day rotation period, this leads to strong degeneracies between the transit and baseline models, and difficulty in achieving an adequate fit to the transit. Therefore, we turn such effects off by setting the rotation period to 100$\times$ its actual value. V1298~Tau~c's transit duration (4.78~hr) is only a modest fraction of the stellar rotation period ($\sim$6\%), and the star could not have rotated by more than approximately 23$^{\circ}$ between the beginning and end of the transit. This is insufficient for any of the occulted spots to have rotated into or out of view during transit, nor moved significantly while being occulted. Stellar rotation would also not significantly affect the apparent spot contrasts. Therefore, we expect little, if not negligible, systematic error in our results due to this assumption. 

We also assumed we are looking directly edge-on at the stellar rotation axis (i.e., a stellar inclination $i_\star$ of 90$^{\circ}$). This is consistent with previous estimates that $i_\star$ = 85.1$^{\circ} \pm 3.6^{\circ}$ \citep{johnson2022}. We present an in-depth analysis of the SCEs themselves in \cite{murphy2026_spots}.

\subsection{Test Case 4: Modeling Spot Crossing Events using Gaussian Processes} \label{subsec:lcfitting_case4gp}

Finally, as a fully flexible and agnostic approach, we utilized a Gaussian Process (GP) model. Specifically, we used \texttt{celerite} \citep{celerite} with a ``Real Term'' or exponential decay kernel $k\left(\tau\right)$ of the form
\begin{equation}
    k\left(\tau\right) = a_{GP} \exp\left(-c_{GP}~\tau\right). \label{eqn:GPterm}
\end{equation}
We chose this kernel due to its simplicity to fit, and ability to represent the morphology of features like SCEs. With this kernel, $\log\left(a_{GP}\right)$ and $\log\left(c_{GP}\right)$ are inputs to the GP model, which is calculated on the residuals between the data and the product of the baseline and GP-free transit model. The GP's relative flux contribution $F_{\mathrm{GP}}$ is additive with the transit and baseline models. Therefore, the full light curve model is
\begin{equation}
    F_{\mathrm{obs, 4}} \left(t\right) = \left( T\left(\theta, t\right) \times S\left(\theta, t\right) \right) + F_{\mathrm{GP}}\,. \label{eqn:case4model}
\end{equation}
We again used \texttt{batman} for the transit model $T\left(\theta, t\right)$. 

To mitigate overflexibility, we freely fit $\log\left(a_{GP}\right)$ and $\log\left(c_{GP}\right)$ on the broadband light curve, then left both values fixed when fitting the spectroscopic light curves. 

\subsection{Limb Darkening Choice} \label{subsec:lcfitting_limbdarkening}

Recent work suggests that quadratic limb darkening laws, used in a majority of exoplanet transit analyses to date, may not accurately represent the intensity profile of stellar disks. At wavelengths where limb darkening is significant, this can lead to biased inferences from the light curve fitting \citep[e.g.,][]{espinoza2016_limbdarkening, keers2024, mercier2025_limbdarkening}. Furthermore, limb darkening profiles can vary significantly between stars of the same spectral type depending on their magnetic properties \citep{kostogryz2024_limbdarkening}. Stellar magnetic properties are typically unknown and not adjustable in the tools used to compute model limb darkening coefficients.

There has been recent momentum toward considering more complex limb darkening when fitting transit light curves, such as non-linear profiles \citep[e.g.,][]{thao2024, ahrer2025_kelt7b, ahrer2025_wasp94b}. To explore this issue in our data, we tested three limb darkening parameterizations when fitting the broadband light curve---quadratic, logarithmic, and four-term non-linear---as well as using model-derived Gaussian priors on the coefficients versus only uninformed uniform priors of [0,1]. For the Gaussian priors, we used values computed for V1298~Tau from the ATLAS stellar model \citep{atlas} via the online \texttt{exoctk} tool\footnote{\url{https://exoctk.stsci.edu/limb_darkening}} \citep{exoctk}. We refit the broadband light curve with SCEs masked (Case 1), using each of these parameterizations and each of the baseline polynomial orders (2--5) mentioned in the previous section. All together, we tested 24 different combinations of parameters and compared the resulting BIC and $\log$ likelihood values, calculated at the median of each parameter's posterior distribution. These are listed in Table~\ref{tab:bics}.

We found that the fits using the cubic polynomial were always preferred, validating our choice of baseline model. Between the limb darkening parameterizations, the logarithmic law had a higher likelihood value than both other cases, and was preferred by the BIC over the non-linear law which has two additional parameters. Additionally, we find the uniform prior fits had slightly higher likelihood values than the Gaussian prior fits, though not significantly enough to determine whether the modeled coefficient values are inaccurate for this target. Recent works have found that offsetting or scaling stellar model derived coefficient spectra can yield better fits though \citep[e.g.,][]{rustamkulov2023_wasp39bERS, sing2024_wasp107, schlawin2024_gj1214}.

We compare our freely-fitted coefficients to model values in Appendix~\ref{apx:LDprofiles}. Based on these tests, we adopted the logarithmic limb darkening law with freely-fit coefficient values in all of our fits except for with \texttt{fleck} (Case 3), as \texttt{fleck} does not support a logarithmic profile, in which case we reverted to a quadratic profile. In any case, this choice did not significantly affect our results.

\section{Broadband Fit Results} \label{sec:bbfits}

\begin{figure*}
    \centering
    \includegraphics[width=\linewidth]{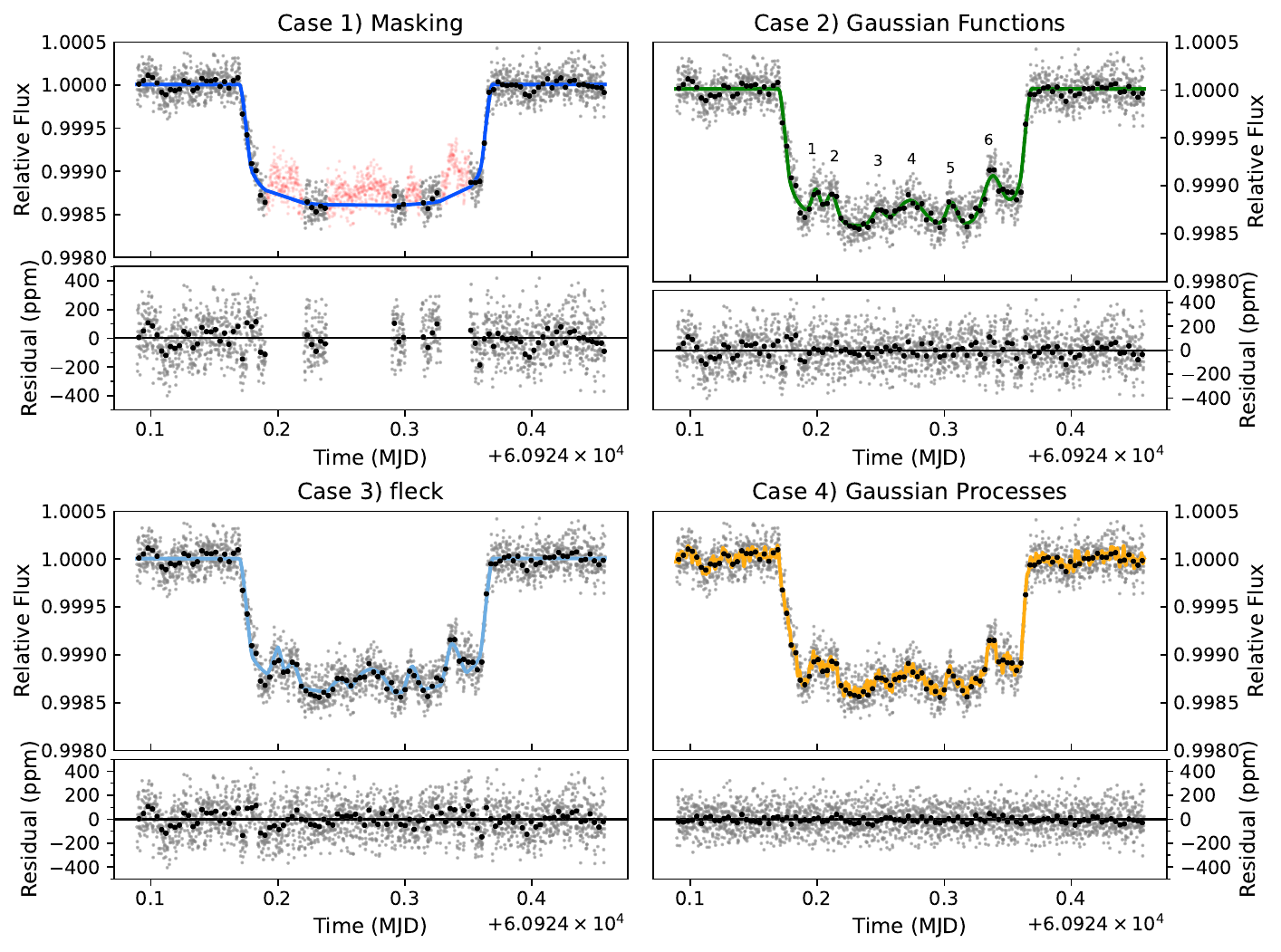}
    \caption{Detrended light curves and best-fit models of the NIRISS/SOSS Order 1 broadband light curve. Gray points are at the native time cadence of $\sim$13~s, and black points are binned to $\sim$5~min cadence. Each panel shows the result of the four methods we implemented to account for stellar surface heterogeneity features in the light curve: 1) masking the features, 2) modeling them as Gaussians, 3) modeling them with \texttt{fleck}, and 4) modeling them with a GP.
    \href{https://github.com/kronos-jwst/KRONOS-I-V1298-Tau-c-NIRISS/blob/main/Create_Figure2.ipynb}{\githubicon}
    }
    \label{fig:bbfits}
\end{figure*}

We achieved good fits to the broadband NIRISS/SOSS light curve in each of the four fitting cases. Figure~\ref{fig:bbfits} shows the detrended light curves, best-fit astrophysical models, and residuals. Each set of residuals are Gaussian distributed with standard deviations of 139~ppm, 138~ppm, 138~ppm, and 104~ppm, respectively. 

To ensure our models are not overfitting the observations, we ran an additional fit treating the transit as if it were completely spot-free. This effectively repeats our Case 1 fit but with no in-transit masks. In this spot-free fit, the out-of-transit residuals were similarly Gaussian distributed with a standard deviation of 137~ppm, compared to 137, 136, 137, and 105~ppm for our nominal Cases~1--4, respectively. However, the in-transit residuals were higher by a factor of 1.25 in the spot-free fit, with a standard deviation of 170~ppm. On the other hand, the in-transit residuals for Cases~1--4 were 144~ppm, 138~ppm, 140~ppm, and 103~ppm, respectively; the in- and out-of-transit residuals in Cases 1--4 are consistent. This demonstrates that neglecting the in-transit SCEs indeed yields a poor fit, and motivates our different treatment methods in the light curve modeling.

\begin{table*}
    \centering
    \caption{Key best-fit and derived orbital, planetary, and systematic parameters from our fits to the NIRISS/SOSS Order 1 broadband light curve.}
    \begin{tabular}{c|c|c|c|c|c} \hline 
    Parameter & Units & Case 1 & Case 2 (fiducial) & Case 3 & Case 4 \\ \hline 
    \multicolumn{6}{c}{\textit{Astrophysical}} \\ \hline 
        $t_\mathrm{c}$ - 2460924                    & BJD$_{\text{TDB}}$ & 0.7691 $\pm$ 0.0001 & 0.7690 $\pm$ 0.0001 & 0.7691 $\pm$ 0.0001 & 0.7691 $\pm$ 0.0002 \\
        $\log_{10} \left(a/R_\star\right)$ &          & 1.083 $\pm$ 0.014   & 1.087 $\pm$ 0.014   & 1.062 $\pm$ 0.010   & 1.069 $\pm$ 0.011 \\
        $a/R_\star$                        &               & 12.11 $\pm$ 0.38    & 12.21 $\pm$ 0.39    & 11.54 $\pm$ 0.28    & 11.73 $\pm$ 0.30 \\
        $\cos\left(i\right)$               &          & 0.039 $\pm$ 0.006   & 0.038 $\pm$ 0.006   & 0.049 $\pm$ 0.004   & 0.046 $\pm$ 0.004 \\
        $i$                                & deg                & 87.11 $\pm$ 0.33    & 87.80 $\pm$ 0.34    & 87.22 $\pm$ 0.22    & 87.37 $\pm$ 0.23 \\ 
        $b$                                &                          & 0.48 $\pm$ 0.05     & 0.47 $\pm$ 0.06     & 0.56 $\pm$ 0.03     & 0.54 $\pm$ 0.03  \\
        $R_{\mathrm{p}}/R_\star$                      &                  & 0.0357 $\pm$ 0.0003 & 0.0360 $\pm$ 0.0003 & 0.0356 $\pm$ 0.0003 & 0.0347 $\pm$ 0.0002\\
        $\delta$                           & ppm                      & 1279 $\pm$ 19       & 1298 $\pm$ 21       & 1268 $\pm$ 24       & 1206 $\pm$ 16       \\
        $t_{14}$                           & hr                    & 4.773               &  4.777              &   4.784             & 4.758                     \\
 \hline 
    \multicolumn{6}{c}{\textit{Systematic}} \\ \hline 
        $a_3$                              & Cubic Term                    & -0.025 $\pm$ 0.004     & -0.022 $\pm$ 0.004   & -0.0290 $\pm$ 0.004  & -0.034 $\pm$ 0.004   \\
        $a_2$                              & Quadratic Term                & -0.0315 $\pm$ 0.0006   & -0.0314 $\pm$ 0.0006 & -0.0304 $\pm$ 0.0006 & -0.0318 $\pm$ 0.0006  \\
        $a_1$                              & Linear Term                   & -0.03863 $\pm$ 0.0001& -0.0387 $\pm$ 0.0001 & -0.0376 $\pm$ 0.0001 & -0.0384 $\pm$ 0.0001 \\
        $a_0$                              & Constant Term                 & 1.00132 $\pm$ 0.00001  & 1.00132 $\pm$ 0.00001& 1.00193 $\pm$ 0.00002& 1.00132 $\pm$ 0.00001   \\ \hline 
        
    \end{tabular}
    \tablecomments{``Best-fit" refers to the median of the posterior distribution, and the given 1$\sigma$ uncertainties are the 16th and 84th percentiles. Derived parameters not defined in the text: $b$ is the impact parameter, $\delta$ is the transit depth, and $t_{14}$ is the transit duration. The $t_{14}$ values are estimated from the median transit models.}
    \label{tab:bbfit_params}
\end{table*}

Table~\ref{tab:bbfit_params} lists the best-fit and derived parameters from each fitting case. The orbital parameters ($t_\mathrm{c}$, $a/R_\star$, $i$) are all consistent within 1.5$\sigma$. Each $a/R_\star$ value is mildly discrepant with the previous measurement of 13.19 $\pm$ 0.55 by \cite{david2019_v1298tau} based on K2 transits, by 1.6$\sigma$, 1.5$\sigma$, 2.7$\sigma$, and 2.3$\sigma$, respectively. The $i$ values are slightly closer to \cite{david2019_v1298tau}'s value of 88.49$^{+0.92}_{-0.72}$~deg, differing by 1.7$\sigma$, 0.87$\sigma$, 1.7$\sigma$, and 1.5$\sigma$, respectively. 

As expected, $a/R_\star$ and $i$ are correlated in each fit. Additionally, the exact orbital solution is influenced by the SCE fitting method used. For example, the spot parameters exhibit no obvious correlation with $a/R_\star$ or $i$ in the Gaussian (Case 2) fits. By contrast, in the \texttt{fleck} (Case 3) fits, the spot latitude parameters are each positively correlated with both $a/R_\star$ and $i$. Therefore, the values for these parameters may be slightly biased. Combining these data with a future transit of V1298~Tau~c with JWST/NIRSpec G395H ($\sim$2.8--5~$\mu$m) will help discern the correct solution. Fortunately, these varied orbital solutions only lead to small vertical offsets between the corresponding transmission spectra, and do not affect its shape nor the atmospheric parameters we derive.

From the Gaussian and \texttt{fleck} fits (Cases 2 and 3), we find that the transit is best fit including six distinct SCEs. These are labeled in the upper-right panel of Figure~\ref{fig:bbfits}. The first two SCEs are very close to one another, so we tested treating this feature as a single SCE. However, treating them individually provided a better fit in terms of the likelihood, and was preferred by the BIC. We repeated this test with SCEs 3 and 4 as well. We verified that our choice in the number of SCEs did not significantly affect the resulting transmission spectrum.

Between all four fits, the GP (Case 4) model results in the highest likelihood value as calculated using the posterior-median parameter values. However, this is more representative of its higher flexibility at the individual integration level rather than being a ``better'' fit. Between Cases 1--3, the Gaussian (Case 2) model has the highest likelihood value and the lowest (i.e., most preferred) BIC value. Methodologically, we also believe Case 2 strikes an optimal balance between agnosticism and flexibility in accounting for the SCEs. For these reasons, we adopt Case 2 as our fiducial case.

Given the prevalence of starspots throughout the transit chord, it is reasonable to expect additional starspots on the stellar limb. However, there are no visible SCEs during ingress nor egress, though such spots would naturally cause lower amplitude light curve features \citep{oshagh2013}; limb SCEs may be below our noise floor if present. In Case 2, we find an elevated standard deviation of residuals in ingress-only and egress-only (171~ppm and 154~ppm, respectively) regions compared to the full transit (131~ppm). This could be caused by additional small SCEs, but is also degenerate with limb darkening \citep{murray2025_SCEeffects}. It may also be influenced by the limited number of integrations during ingress and egress. Nevertheless, we later demonstrate that our results are robust to variations in limb darkening treatments (Figures~\ref{fig:transmissionspectra} and \ref{fig:LDCcomp}--\ref{fig:LDCcomp_quad}) and spot-induced chromatic transit time variations \citep[][Figure~\ref{fig:limbasymmetry} in this work]{oshagh2013}. Therefore, even if spots are present on the stellar limb, they have a negligible impact on our results.

%===========================================================%
\section{The NIRISS/SOSS Transmission Spectrum of V1298 Tau c}\label{sec:spectrum}

\begin{figure*}
    \centering
    \includegraphics[width=\linewidth]{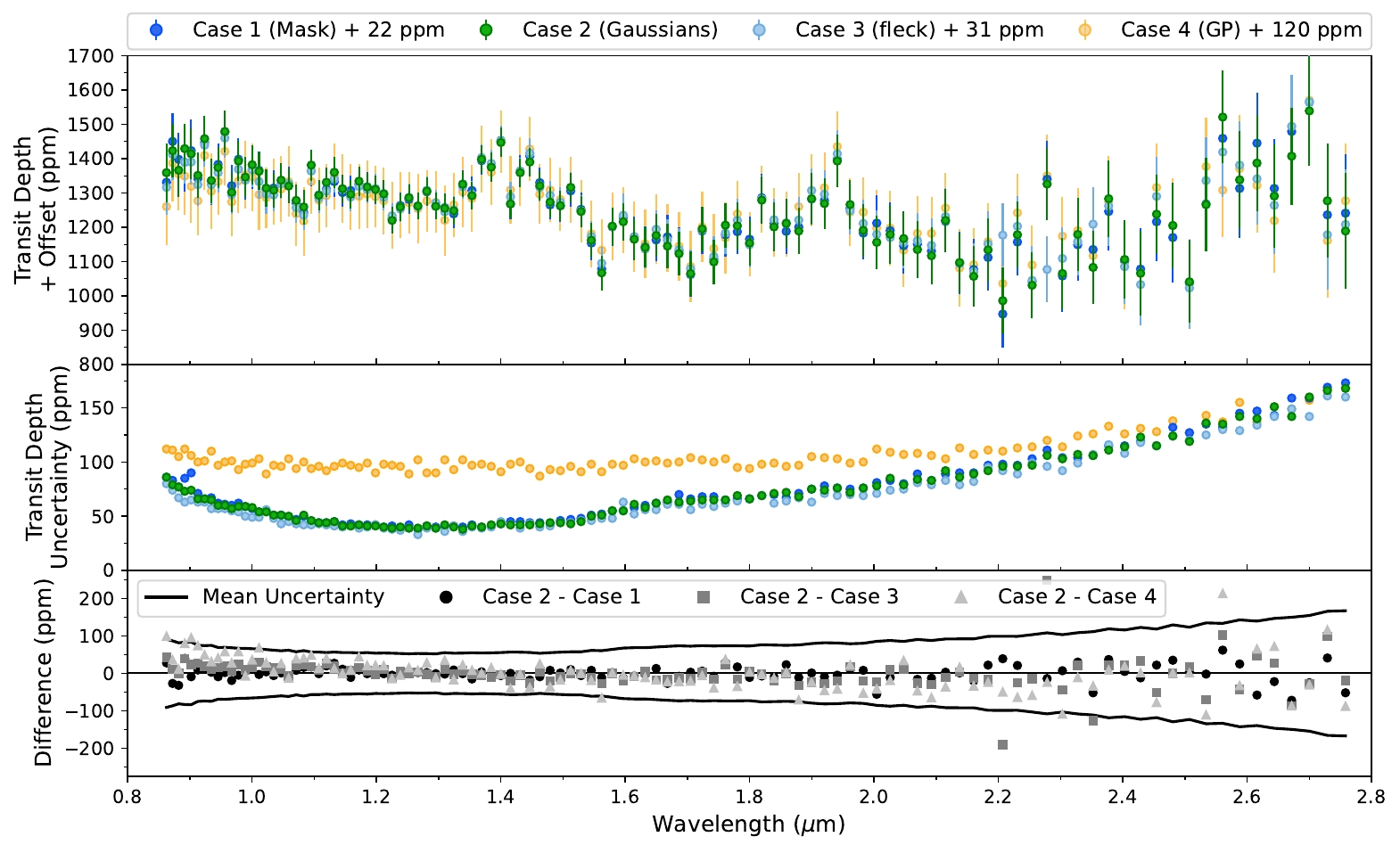}
    \caption{
    The results of four independent methodologies for handling starspot crossing events during a transit of V1298~Tau~c observed with JWST NIRISS/SOSS. Top: The $\sim$0.8--2.8\,$\mu$m transmission spectra of V1298~Tau~c at a resolution of R=100. The color corresponds to the fitting routine used (see Sections~\ref{subsec:lcfitting_case1mask}--\ref{subsec:lcfitting_case4gp}). The spectra are manually offset by 21~ppm for Case 1, 20~ppm for Case 3, and 120~ppm for Case 4 to have the same median transit depth for visual comparison. 
    Middle: The transit depth uncertainties achieved in each method.
    Bottom: The chromatic difference between each case including offsets to highlight the relative shape of each result. The solid lines represent the mean transit depth uncertainty between each case. This figure highlights that regardless of how the observed SCEs are masked or fit, the resulting transmission spectrum has the same overall shape and absorption features.
    \href{https://github.com/kronos-jwst/KRONOS-I-V1298-Tau-c-NIRISS/blob/main/Create_Figure3.ipynb}{\githubicon}
    }
    \label{fig:transmissionspectra}
\end{figure*}

We derived the transmission spectrum using each of the four SCE-handling methods described in Section~\ref{sec:methods}. Figure~\ref{fig:transmissionspectra} presents all four at $R$=100. The middle panel of Figure~\ref{fig:transmissionspectra} shows the transit depth uncertainties in each case. We achieved nearly identical uncertainties between Cases 1--3, ranging from a minimum of $\sim$40~ppm around 1.2~$\mu$m to $\sim$150~ppm at the red-edge of the bandpass. On the other hand, the GP fit (Case 4) uncertainties were significantly larger owing to its flexibility, and hit a floor around $\sim$100~ppm. 

Due to the differences in assumed $a/R_\star$ and $i$ based on the broadband fits and assumptions about the SCEs, there were slight vertical offsets between the derived spectrum for each case. The difference in median transit depth of each spectrum relative to that of Gaussian fits (Case 2) was 21~ppm for the masked fits (Case 1), 30~ppm for \texttt{fleck} (Case 3), and 120~ppm for the GP fits (Case 4). Additionally, we noticed that the mean value of Case 4 for points in-transit between SCEs varies from approximately -5 to -20\,ppm. For both points out-of-transit as well as all in-transit points, this mean value is zero. Therefore, the GP model may be absorbing some of the transit depth, contributing to the higher relative offset.
The spectra shown in Figure~\ref{fig:transmissionspectra} are offset to have equal median transit depths, using Case 2 as the reference point, to compare their relative shapes. We find that the shape and relative feature amplitudes in each spectrum are consistent between the different SCE fitting methods, raising our confidence that the specific method should not impact our inference of V1298~Tau~c's atmospheric properties. 

To further highlight the consistency between spectra, the lower panel of Figure~\ref{fig:transmissionspectra} shows the differences between the spectra once these median differences have been subtracted, compared to the mean transit depth uncertainty (across all cases) as a function of wavelength. In nearly every channel, the relative difference between each spectrum is much less than the uncertainty. The masked (Case 1) and Gaussian (Case 2) results are particularly consistent. The largest discrepancy is with the GP fit (Case 4) which appears to have a slight slope, particularly at the shortest wavelengths. However, this minor slope is still within the magnitude of uncertainties. 

\section{The NIRISS/SOSS Stellar Spectrum of V1298~Tau} \label{sec:stellarspectrum}

Exoplanet transmission spectra, particularly at the relatively short wavelengths of NIRISS/SOSS, are subject to contamination due to the transit light source effect \citep[TLSE; ][]{rackham2018_tlseM, rackham2019_tlseFGK}. This is especially a concern given the numerous starspot crossings during our observation, and the generally high activity level of V1298~Tau \citep{david2019_v1298tau, feinstein2021_v1298tauGemini, suarez_mascareno_2022, feinstein2022_v1298tauTESS}. To account for potential contamination, we first fit the stellar spectrum measured out of transit to estimate the properties of surface heterogeneities on V1298~Tau. These will be used as priors when fitting and interpreting our transmission spectrum. 

\subsection{Stellar Spectrum Fit Setup} \label{subsec:stellarspectrum_setup}
% - out of transit stellar spectrum fits:
We constructed the stellar spectrum as the median of all out-of-transit flux-calibrated 1D spectra. We conservatively estimated the transit contact points from the broadband light curve, defining integrations 1--521 as pre-transit and 1800--2391 as post-transit, and clipping the possible mirror-tilt event near the end of the observation. We propagated the uncertainty as the standard error on the median. 

We fit the out-of-transit spectrum using an interpolated grid of model stellar spectra drawn from the NewEra grid \citep{NewEraGrid} using \texttt{speclib} \citep{Rackham2023, Rackham2024}, following the approach of previous HST and JWST analyses \citep{Davoudi2024, Narrett2024, Rathcke2025, Niraula2026}. We assume a fixed $\log\left(\mathrm{g} / \left[\mathrm{cm}/\mathrm{s}^2\right]\right)$ = 4.25, based on literature estimates ranging from 4.227 \citep{finociety_2023} to 4.246 \citep{david19_b}, and interpolate to this value from the pre-calculated grid models at $log\left(g\right)$ = 4.0 and 4.5. We also assume a fixed [Fe/H] = 0.1 \citep{suarez_mascareno_2022}. 

To account for surface heterogeneities, we assembled the spectrum from up to three components. The first component is the quiet photosphere, and the additional components represent cooler contributions (e.g., from spots) and hotter contributions (e.g., from faculae). Including more than these three components led to significant degeneracies. We prescribed each component an effective temperature and covering fraction as free parameters\footnote{For clarity: the photosphere's covering fraction is not a free parameter, and is always assumed to be the remainder of unity and the sum of the other components' covering fractions.}. We then scaled the combined model spectrum to the flux received at Earth for comparison with our observation, using the stellar radius $R_\star$ as a free parameter but a fixed distance between V1298~Tau and Earth of 108.2~pc \citep{gaia_dr2}. We also fit for extinction in terms of $A_V$, which we converted to a multiplicative wavelength-dependent attenuation factor following \cite{CCM89} when computing the model. All together, we fit for $R_\star$, $A_V$, the photosphere temperature $T_\mathrm{phot}$, and the $T$ and covering fraction $f$ of the cooler and hotter components. We enforced a Gaussian prior on $R_\star$ of 1.278 $\pm$ 0.07 \citep{suarez_mascareno_2022}, and uniform priors of (0, 0.3) on $A_V$, (3000~K, 6000~K) on the temperatures, and (0.0, 0.5) on the covering fractions. Our range of temperatures is consistent with both cooler and hotter regions on the solar photosphere.

We used the nested sampling Monte Carlo algorithm \texttt{MLFriends} \citep{ultranest_algorithm1, ultranest_algorithm2} implemented in the python package \texttt{UltraNest} \citep{ultranest} to derive the respective posterior probability distributions and compare the Bayesian evidence of each model. 
We used 200 live points to balance computational efficiency against adequate exploration of the parameter space, and adopted \texttt{frac\_remain}$= 0.2$ to terminate the runs once the remaining posterior mass is negligible.

\subsection{Stellar Spectrum Fit Results} \label{subsec:stellarspectrum_result}

\begin{figure}
    \centering
    \includegraphics[width=\linewidth]{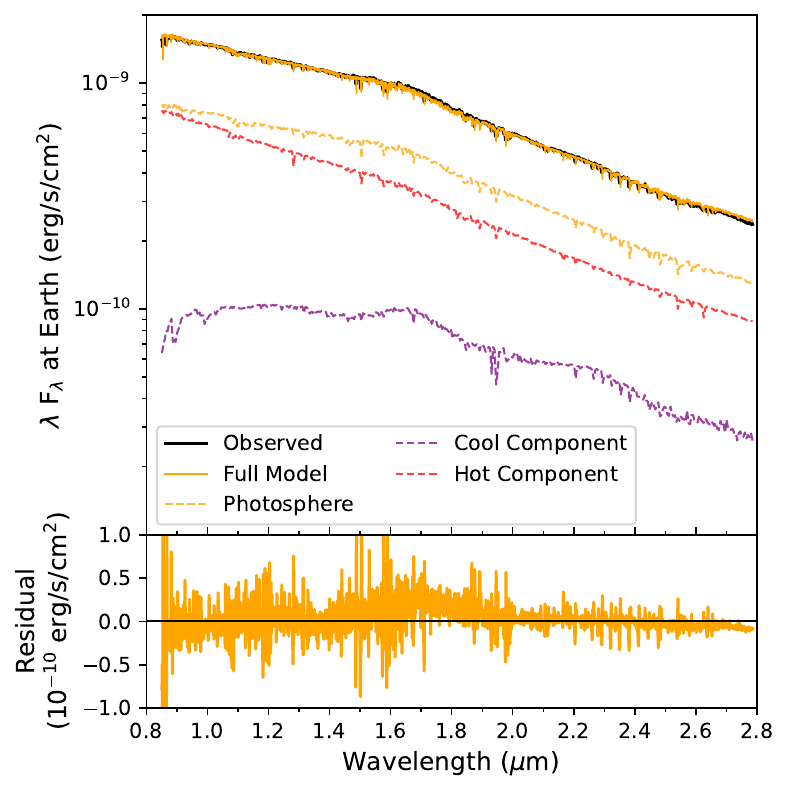}
    \caption{The 0.8--2.8$\mu$m spectrum of V1298~Tau. 
    Top: Comparing the observed spectrum to the best-fit 3-component model, and the individual components scaled by their respective covering fractions. Bottom: The corresponding data-model residuals. NIRISS/SOSS is flux calibrated to the $\sim$1\% level according to the JWST User Documentation which is on the order of the magnitude of our residuals relative to the spectrum ($\lesssim$5\%).
    \href{https://github.com/kronos-jwst/KRONOS-I-V1298-Tau-c-NIRISS/blob/main/Create_Figure4.ipynb}{\githubicon}
    }
    \label{fig:stellarspectrum}
\end{figure}

% \begin{table}
%     \centering
%     \caption{Stellar properties of V1298~Tau derived in this work from fits with free and fixed extinction \citep[based on][]{david19_b}.}
%     \begin{tabular}{c|c|c|c} \hline 
%     Parameter                  & Unit        & Value (Free $A_V$)  & Value ($A_V$ = 0.074)      \\ \hline 
%     $R_\star$                  & $R_\odot$   &  1.292 $\pm$ 0.002  &  1.278 $\pm$ 0.002          \\
%     $T_{\mathrm{phot}}$        & K           &  4900 $\pm$ 7       &  4899 $\pm$ 7               \\   
%     $T_{\mathrm{cool}}$        & K           &  3372 $\pm$ 23      &  3375 $\pm$ 23              \\
%     $f_\mathrm{\mathrm{cool}}$ &             &  0.198 $\pm$ 0.006  &  0.163 $\pm$ 0.007          \\
%     $T_{\mathrm{hot}}$         & K           &  5560 $\pm$ 40      &  5609 $\pm$ 37              \\
%     $f_{\mathrm{hot}}$         &             & 0.32 $\pm$ 0.02     &  0.35 $\pm$ 0.02            \\ \hline 
%     \end{tabular}
%     \label{tab:starproperties}
% \end{table}

%% with new Fe
\begin{table}
    \centering
    \caption{Stellar properties of V1298~Tau derived in this work from fits with free and fixed extinction \citep[based on][]{david19_b}.}
    \begin{tabular}{c|c|c|c} \hline 
    Parameter                  & Unit        & Value (Free $A_V$)  & Value ($A_V$ = 0.074)      \\ \hline 
    $R_\star$                  & $R_\odot$   &  1.300 $\pm$ 0.002  &  1.288 $\pm$ 0.002          \\
    $T_{\mathrm{phot}}$        & K           &  4893 $\pm$ 10       &  4893 $\pm$ 10               \\   
    $T_{\mathrm{cool}}$        & K           &  3400 $\pm$ 5      &  3400 $\pm$ 6              \\
    $f_\mathrm{\mathrm{cool}}$ &             &  0.224 $\pm$ 0.006  &  0.192 $\pm$ 0.007          \\
    $T_{\mathrm{hot}}$         & K           &  5637 $\pm$ 31      &  5667 $\pm$ 27              \\
    $f_{\mathrm{hot}}$         &             & 0.283 $\pm$ 0.02     &  0.31 $\pm$ 0.02            \\ \hline 
    \end{tabular}
    \label{tab:starproperties}
\end{table}

Figure~\ref{fig:stellarspectrum} shows the observed spectrum compared to the best-fit three-component model, as well as the individual components scaled by their covering fractions. We found that a three-component model was preferred in terms of the Bayesian Evidence by $\Delta \log\left(Z\right)$ = 744 ($\sim$39$\sigma$) over the one-component model and $\Delta \log\left(Z\right)$ = 114 ($\sim$15$\sigma$) over the two-component model. The two-component model was also preferred over the one-component model by $\Delta \log\left(Z\right)$ = 630 ($\sim$35$\sigma$).  

Table~\ref{tab:starproperties} lists the best-fit parameter values from the three-component fit. The best-fit stellar radius is $R_\star$ = 1.300 $\pm$ 0.002, slightly higher than our prior but broadly consistent with literature measurements, which range from 1.278 $\pm$ 0.07 \citep{suarez_mascareno_2022} to 1.43 $\pm$ 0.03 \citep{finociety_2023}. We find a photosphere temperature of $T_\mathrm{phot} = 4893 \pm 10$\,K, consistent with literature $T_\mathrm{eff}$ measurements \citep{suarez_mascareno_2022, finociety_2023, livingston2026_v1298tau}. Then, we find the cooler component has $T_\mathrm{cool} = 3400 \pm 5$\,K with a covering fraction $f_\mathrm{cool} = 0.224 \pm 0.006$, and the hotter component has $T_\mathrm{hot} = 5637 \pm 31$\,K with a covering fraction of $f_\mathrm{hot} = 0.28 \pm 0.02$. Together, these results suggest that surface heterogeneities make up a large fraction of V1298~Tau's visible surface.

The temperature of this cool component, which is likely composed of starspots, is broadly consistent with average Sunspot temperatures \citep{Solanki2003} and empirical expectations based on the photosphere temperature \citep{rackham2019_tlseFGK}. The covering fraction of these starspots is twice as high as previously measured for V1298~Tau based on Kepler photometry \citep{Morris2020_spotsVSage}, and is within the wide range of $\sim$10--80\% seen on other young stars \citep{stauffer2003_spottyyoungKdwarfs, GullySantiago2017_youngspottystar, Morris2020_spotsVSage}. We will explore the detailed starspot properties of V1298~Tau further in the follow up work \cite{murphy2026_spots}. The hot component is $\sim$740~K hotter than the photosphere. This is higher than for Solar faculae, which are typically only a few hundred K hotter than the photosphere \citep{wang1998_faculae, sutterlin1999, solovev2019_faculae}, but this component may also include contribution from hotter plages. 
%Nevertheless, if this hot component is primarily composed of faculae, these covering fractions suggest a $\sim$1.25:1 ratio of faculae-to-spot area which is consistent with measurements on the Sun \citep{chapman1997}.

The $A_V$ posterior piled up at zero with a 3$\sigma$ upper limit of $A_V$ = 0.0012. Based on maps of the local interstellar medium, \cite{david19_b} previously estimated a color excess of E($B-V$) = 0.024 $\pm$ 0.015 which, assuming an average extinction $R_V = 3.1$ \citep{Draine2003}, corresponds to $A_V = 0.074 \pm 0.047$. This is consistent with our upper-limit at $\sim$1.5$\sigma$. We tested repeating these fits with $A_V$ fixed to 0.074. The resulting temperatures for each component were highly consistent with the free $A_V$ result ($T_{\mathrm{phot}}$ = 4893 $\pm$ 10\,K, $T_{\mathrm{cool}}$ = 3400 $\pm$ 6\,K, $T_{\mathrm{hot}}$ = 5667 $\pm$ 27\,K). The cool component covering fraction was slightly lower ($f_\mathrm{\mathrm{cool}}$ = 0.192 $\pm$ 0.007), but the hot component covering fraction was consistent ($f_\mathrm{\mathrm{hot}}$ = 0.31 $\pm$ 0.02). Therefore, the exact covering fraction of starspots depends on the assumed extinction, but is generally near 20\%. The stellar radius was also lower ($R_\star$ = 1.288 $\pm$ 0.002 $R_\odot$), which makes sense as this primarily scales the entire spectrum to relatively higher flux, compensating for the higher extinction. However, this radius is still consistent with literature measurements.

\section{Interpreting the Atmosphere of V1298 Tau c} \label{sec:atmomodeling}

We present a suite of atmospheric retrieval models to robustly determine if observed features in the transmission spectrum originate from V1298~Tau~c, and infer its atmospheric composition.

\subsection{Modeling Details} \label{subsec:atmomodeling_retrieval}

To interpret our observed transmission spectrum and infer the atmospheric properties of V1298~Tau~c's atmosphere, we perform inverse modeling using the \texttt{Aurora} framework \citep{aurora}. \texttt{Aurora} combines an atmospheric forward model with Bayesian parameter estimation, and in this work we include the developments presented in \cite{Pinhas2018}, \cite{welbanks2019}, \cite{Nixon2020, Nixon2022}, and \cite{Nixon2024}.

The \texttt{Aurora} model uses the parametric pressure--temperature profile described by \cite{madhu2009}, with 100 layers spaced uniformly in log-pressure space from 10$^{-9}$ to 100~bar. We assume a background gas of H$_2$ and He in Solar proportions \citep{asplund2009}. We fix $R_{\star}=1.292 R_{\odot}$, $\log_{10} g_{\star}=4.25$, consistent with previous analyses in this study, and assume [Fe/H]$_{\star}=0.1$ \citep{suarez_mascareno_2022}. We consider absorption from the following chemical species: H$_2$O \citep{Rothman2010}, CH$_4$ \citep{Yurchenko2014}, NH$_3$, HCN \citep{Barber2014}, \citep{Yurchenko2011}, CO \citep{Rothman2010} and CO$_2$ \citep{Rothman2010}; as well as H$_2$/H$_2$ and H$_2$/He collision-induced absorption \citep{Richard2012}. We account for aerosol opacity using the patchy cloud model described by \cite{Line2016}, as well as stellar heterogeneity effects following \cite{rackham2018_tlseM} and \cite{Pinhas2018}. We also fit for the mass and radius of the planet. Table~\ref{tab:retrievalpriors} lists all free parameters and priors.

\begin{table}[ht]
    \centering
    \caption{Priors used for atmospheric parameter estimation.}
    \begin{tabular}{c|c|c} \hline 
    Parameter & Units & Prior \\ \hline 
    \multicolumn{3}{c}{ \textit{Chemistry and P-T Profile}} \\ \hline 
    $\log_{10}\left( X_i\right)$ & unitless & [-12, -0.3] \\
    $T_0$                          & K        & [300, 1500]  \\
    $\alpha_1$, $\alpha_2$      & K$^{-1/2}$ & [0.02, 2.0] \\
    $\log_{10}\left(P_1, P_2, \mathrm{ and~} P_{\rm ref}\right)$ & bar & [-9, 2] \\
    $\log_{10}\left(P_3\right)$ & bar & [-2, 2] \\ \hline 
    \multicolumn{3}{c}{ \textit{Bulk Parameters}} \\ \hline 
    $M_p$ & $M_{\oplus}$ & [1, 100] \\
    $R_p$ & $R_{\oplus}$ & $\mathcal{N}\left(5.1, 0.1\right)$ \\ \hline 
    \multicolumn{3}{c}{ \textit{Aerosols (Clouds/Hazes)}} \\ \hline 
    $\log_{10} \left(a\right)$ & unitless & [-4, 10] \\
        $\gamma$ & unitless & [-20, 2] \\
        $\log_{10} \left(P_c\right)$ & bar & [-9, 2] \\
        $\phi_{\rm c+h}$ & unitless & [0, 1] \\ \hline 
    \multicolumn{3}{c}{ \textit{Stellar Heterogeneity: Informed Priors}} \\ \hline 
    $f_{\mathrm{het}}$ & unitless & $\mathcal{N}\left(0.20,0.01\right)$ \\
    $T_{\mathrm{het}}$ & K &  $\mathcal{N}\left(3400, 100\right)$ \\
    $T_{\mathrm{phot}}$ & K & $\mathcal{N}\left(4900, 50\right)$ \\ \hline 
    \multicolumn{3}{c}{ \textit{Stellar Heterogeneity: Uninformed Priors}} \\ \hline 
    $f_{\mathrm{het}}$ & unitless & [0, 0.5] \\
    $T_{\mathrm{het}}$ & K & [2450, 7350] \\ 
    $T_{\mathrm{phot}}$ & K & $\mathcal{N}\left(4900, 50\right)$ \\ \hline 
    \end{tabular}
    \tablecomments{ $X_i$ denotes the volume mixing ratio of chemical species $i$, and we enforce that $\sum_i X_i = 1$. $\mathcal{N}$ refers to Gaussian (i.e., normal) priors and brackets refer to uniform priors. Note that we also tested uniform priors on T$_\mathrm{phot}$, T$_\mathrm{het}$, and $f_{\mathrm{het}}$ in addition to the listed Gaussian priors.}
    \label{tab:retrievalpriors}
\end{table}

To fully explore the impact of the TLSE on our retrieved atmospheric parameters, we test both informed and uniformed priors on the stellar heterogeneity parameters: heterogeneity temperature $T_\mathrm{het}$, and heterogeneity covering fraction $f_\mathrm{het}$. While the three-temperature stellar spectrum was preferred, the hotter component should not contribute to changes in the retrieved H$_2$O abundances because it is significantly hotter than the H$_2$O dissociation temperature, so we assume these heterogeneities are starspots. The informed priors take the form of Normal distributions based on our fits to the stellar spectrum, but with slightly more conservative uncertainties. The uninformed priors take the form of wide uniform distributions and provide an independent check on these stellar heterogeneity properties. In both cases we use the same prior on the photosphere temperature $T_{\rm phot}$. Comparing both results reveals the sensitivity of our inferred atmospheric parameters to the accuracy of these values. Both sets of priors are listed in Table~\ref{tab:retrievalpriors}. We discuss the impact of the TLSE further in Section~\ref{subsec:retrievals_TLSE}.

We generate high-resolution ($R$=60,000) spectra from 0.83-2.8~$\mu$m using opacity sampling, which are convolved with the point spread function of NIRISS/SOSS and binned to the highest resolution of the observed data ($R$=300). The binned model is used to calculate the likelihood function, assuming each measurement of the transit depth is independent and follows a Gaussian distribution. We explore the model parameter space using the Nested Sampling algorithm \citep{Skilling2006}, specifically \texttt{PyMultiNest} \citep{Buchner2014_pymultinest}, a \texttt{python} interface for MultiNest \citep{Feroz2008, Feroz2009}, employing 1000 live points. Note that, while we primarily fit models to the fiducial Gaussian (Case 2) $R$=300 spectrum, we find consistent results when fitting to other cases and at different resolutions.

\subsection{Atmospheric Properties of V1298~Tau~c} \label{subsec:atmomodeling_results}

\begin{figure*}
    \centering
    \includegraphics[width=\linewidth]{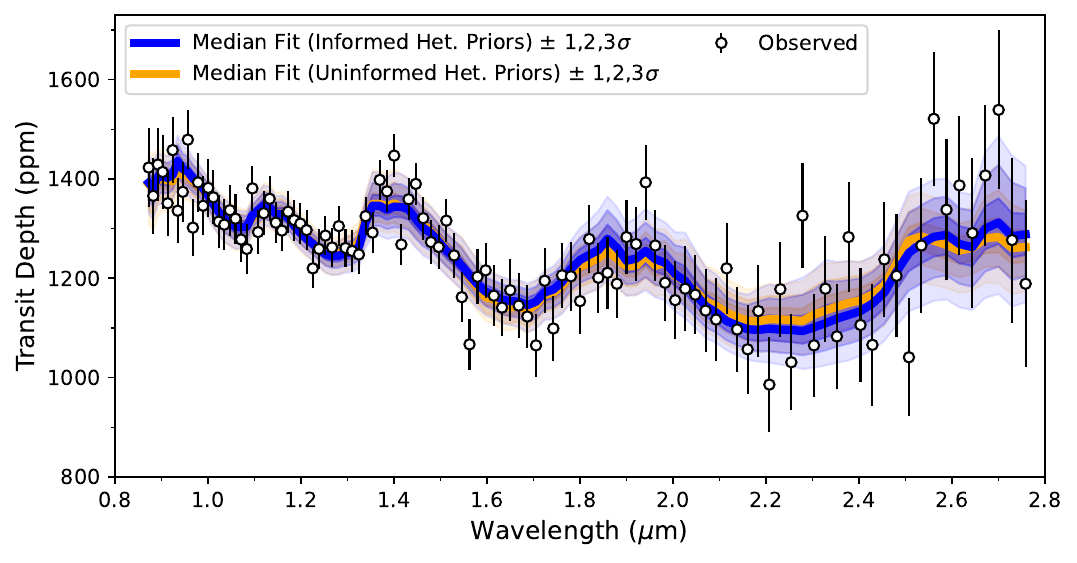}
    \caption{
    Our fiducial transmission spectrum of V1298~Tau~c (black points) compared to the median retrieved models with (blue line) and without (orange line) enforcing informed stellar heterogeneity priors, each binned down to $R$=100. The 1, 2, and 3$\sigma$ model uncertainties are illustrated by the colored shaded regions. We achieve excellent fits to the observed spectrum, which are largely insensitive to the exact stellar heterogeneity parameters. 
     \href{https://github.com/kronos-jwst/KRONOS-I-V1298-Tau-c-NIRISS/blob/main/Create_Figure5.ipynb}{\githubicon}
    }
    \label{fig:retrievals_spectrum}
\end{figure*}

\begin{figure}
    \centering
    \includegraphics[width=1\linewidth]{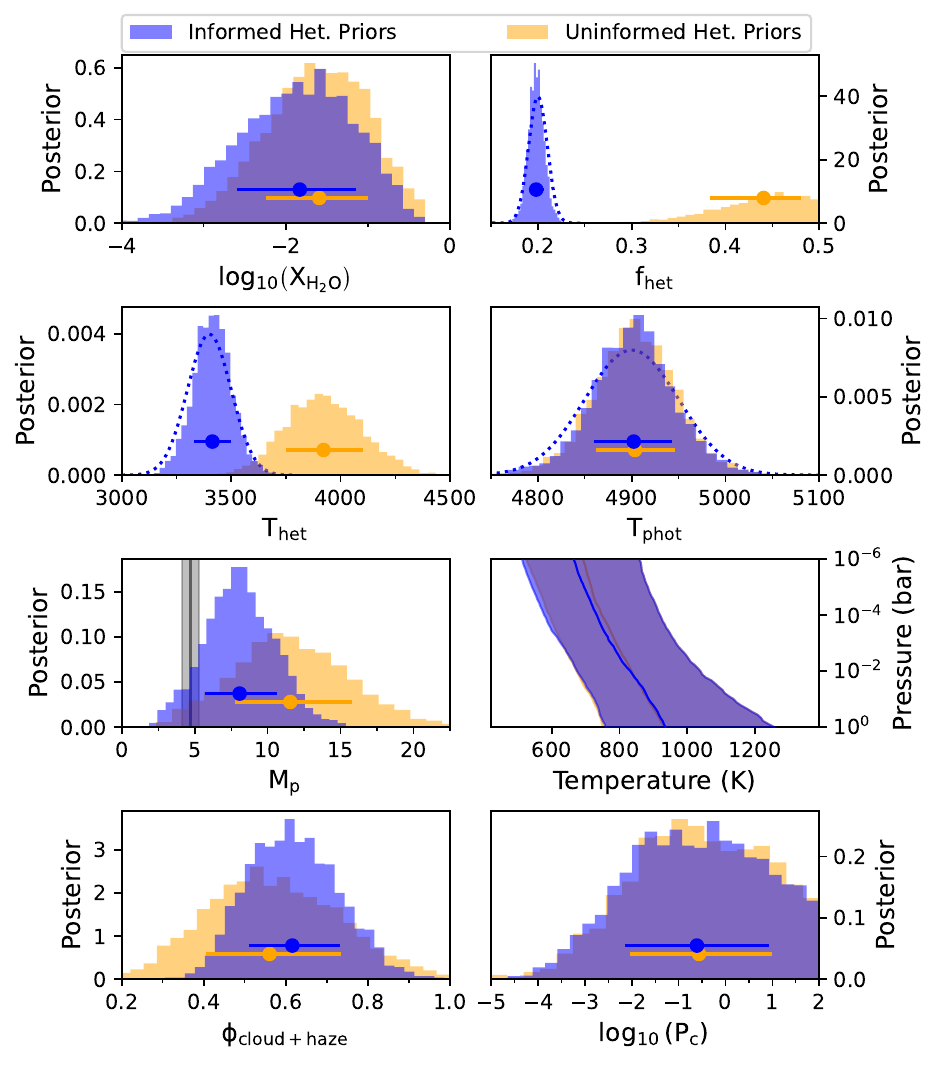}
    \caption{Posterior distributions of key parameters from our atmospheric retrievals with (blue) and without (orange) informed stellar heterogeneity priors, as well as the best-fit pressure-temperature profile. The blue dotted lines illustrate the normal priors applied to $f_{\mathrm{het}}$, $T_\mathrm{het}$, and $T_{\mathrm{phot}}$ for the run with informed priors. The black line in the M$_p$ panel represents the dynamical mass constraint of \cite{livingston2026_v1298tau} for comparison.
    \href{https://github.com/kronos-jwst/KRONOS-I-V1298-Tau-c-NIRISS/blob/main/Create_Figure6.ipynb}{\githubicon}
    }
    \label{fig:retrievals_abundances}
\end{figure}

We present our two median retrieved spectra using informed and uninformed stellar heterogeneity priors in Figure~\ref{fig:retrievals_spectrum}. These are binned to $R$=100 for visual clarity. We find that our model is able to explain the data very well with a best-fit $\chi^2_\nu$ of 0.92 ($\chi^2$=231.60, 275 data points, 22 model parameters). Figure~\ref{fig:retrievals_abundances} shows the posterior distributions for several key parameters: the water abundance, stellar heterogeneity parameters, planet mass, and aerosol parameters; as well as the pressure-temperature profile. The full list of best-fit values is given in Table~\ref{tab:retrievals_vals}, and we provide the full posterior corner plots in the Appendix Section~\ref{apx:retrievalresults}. 

\begin{table}[]
    \centering
     \caption{Atmospheric properties of V1298~Tau~c derived in this work from atmospheric retrievals with informed stellar heterogeneity priors (middle) versus uniformed priors (right).}
    \begin{tabular}{c|c|c|c}
    Parameter & Units & Value (Informed) & (Uninformed) \\ \hline 
\multicolumn{4}{c}{ \textit{Chemistry}} \\ \hline 
$\mathrm{log}_{10}(X_{\rm H_2O})$ & & -1.83$^{+0.68}_{-0.77}$  &  -1.59$^{+0.60}_{-0.65}$ \\
$\mathrm{log}_{10}(X_{\rm CH_4})$ & & -8.38$^{+2.36}_{-2.20}$  &  -8.45$^{+2.22}_{-2.24}$ \\
$\mathrm{log}_{10}(X_{\rm NH_3})$ & & -8.30$^{+2.48}_{-2.28}$  &  -8.43$^{+2.26}_{-2.20}$ \\
$\mathrm{log}_{10}(X_{\rm HCN})$ & & -8.01$^{+2.60}_{-2.47}$  &  -7.83$^{+2.57}_{-2.54}$ \\
$\mathrm{log}_{10}(X_{\rm CO})$ & & -7.05$^{+3.08}_{-2.97}$  &  -7.15$^{+3.09}_{-2.97}$ \\
$\mathrm{log}_{10}(X_{\rm CO_2})$ & & -7.35$^{+2.81}_{-2.89}$  &  -7.12$^{+2.96}_{-2.90}$ \\
\hline \multicolumn{4}{c}{ \textit{P-T Profile}} \\ \hline 
$T_{0}$  & K & 625$^{+181}_{-153}$  &  650$^{+176}_{-158}$ \\
$\alpha_{1}$ & K$^{-1/2}$& 1.30$^{+0.42}_{-0.43}$  &  1.31$^{+0.41}_{-0.43}$ \\
$\alpha_{2}$ & K$^{-1/2}$& 1.08$^{+0.55}_{-0.53}$  &  1.12$^{+0.53}_{-0.52}$ \\
$\mathrm{log}_{10}(P_1)$ & bar & -2.50$^{+1.88}_{-2.33}$  &  -2.35$^{+1.91}_{-2.25}$ \\
$\mathrm{log}_{10}(P_2)$ & bar& -6.04$^{+2.30}_{-1.88}$  &  -6.07$^{+2.48}_{-1.85}$ \\
$\mathrm{log}_{10}(P_3)$ & bar& 0.41$^{+1.05}_{-1.19}$  &  0.53$^{+0.96}_{-1.24}$ \\
$\mathrm{log}_{10}(P_{\rm ref})$ & bar& -4.55$^{+0.68}_{-0.67}$  &  -5.63$^{+0.99}_{-1.16}$ \\
\hline \multicolumn{4}{c}{ \textit{Aerosols}} \\ \hline 
$\mathrm{log}_{10}(a)$ & & 7.37$^{+1.56}_{-1.64}$  &  6.84$^{+1.79}_{-2.01}$ \\
$\gamma$ & & -8.80$^{+2.18}_{-2.35}$  &  -9.64$^{+3.30}_{-3.43}$ \\
$\mathrm{log}_{10}(P_c)$ & & -0.61$^{+1.54}_{-1.53}$  &  -0.56$^{+1.56}_{-1.47}$ \\
$\phi_{\rm c+h}$ & & 0.62$^{+0.12}_{-0.11}$  &  0.56$^{+0.18}_{-0.15}$ \\
\hline \multicolumn{4}{c}{ \textit{Stellar Heterogeneity}} \\ \hline 
$f_{\mathrm{het}}$ & & 0.20$\pm$0.01 &  0.44$^{+0.04}_{-0.06}$ \\
$T_{\rm het}$ & K& 3414$^{+85}_{-87}$  &  3922$^{+181}_{-171}$ \\
$T_{\rm phot}$ & K& 4902$^{+41}_{-42}$  &  4904 $\pm$ 42 \\
\hline \multicolumn{4}{c}{ \textit{Bulk Planetary Properties}} \\ \hline 
$R_p$  & R$_\oplus$ & 5.04$^{+0.09}_{-0.08}$  &  5.03$^{+0.09}_{-0.08}$ \\
$M_p$ & M$_\oplus$ & 8.08$^{+2.54}_{-2.35}$  &  11.55$^{+4.22}_{-3.79}$ 
    \end{tabular}
    \label{tab:retrievals_vals}
\end{table}

We find strong evidence for water in the atmosphere of V1298~Tau~c. For the models with informed stellar heterogeneity priors, a model including water is preferred to one without water with $\Delta \ln \mathcal{Z}=29.30$. For models with uninformed stellar heterogeneity priors, the same comparison yields $\Delta \ln \mathcal{Z}=27.25$. Note that thresholds for ``strong'' or ``decisive'' evidence typically require $\Delta \ln \mathcal{Z}=4.6-5$ \citep{Kass1995,Trotta2008}. We derive the water abundance to be $\log_{10}\left(X_{H_2O}\right)$~=~-1.83$^{+0.68}_{-0.77}$ from the informed prior fit. Based on this, we estimate V1298~Tau~c's atmospheric metallicity in terms of the oxygen abundance to be $\log_{10}\left(\mathrm{O/H}\right) = 1.17^{+0.68}_{-0.77}$, or $14.8^{+56.0}_{-12.28}$ $\times$ solar. 

In addition to H$_2$O, we find V1298~Tau~c has a deeper transit depth in the He~\textrm{I} metastable triplet at 1.083~$\mu$m and Paschen~$\alpha$ at 1.875~$\mu$m. This excess absorption may be due to an escaping hydrogen/helium envelope. Detailed discussion and modeling of this potential escape signature is presented in Barat et al. (in prep).

We do not find significant evidence for any additional molecules. Methane and hydrogen cyanide also have prominent absorption features at these wavelengths, but the abundances of these species are not constrained by the present data. As a result, we are unable to constrain the atmospheric carbon-to-oxygen ratio. These non-detections may be due in part to the overlapping of absorption bands and the relative strength of water absorption at these wavelengths. Combined with the spectrum presented here, our upcoming transit observation with JWST NIRSpec/G395H ($\sim$3--5\,$\mu$m) will provide more insights into the full chemical inventory of V1298~Tau~c.

We note that there are slight differences in posterior distributions of the stellar heterogeneity properties between the informed and uninformed models. The posteriors for $f_{\rm het}$, $T_{\rm het}$, and $T_{\rm phot}$ from the informed priors model are consistent with their priors, and likely prior-dominated. On the other hand, the posteriors from the uninformed priors model suggest a higher $T_{\rm het}$ and $f_{\rm het}$ (Table~\ref{tab:retrievals_vals}). Although we do not see significant correlation between $f_\mathrm{het}$ and $T_{\mathrm{het}}$ (Figure~\ref{fig:cornerplot_uninformedpriors}) in the uninformed case, the lower temperature contrast relative to the photosphere could be compensated by the higher covering fraction to produce the same TLSE slope. A 44\% covering fraction of spots could be plausible for a young and highly active star \citep[e.g.,][]{Grankin1999_youngspottystar, stauffer2003_spottyyoungKdwarfs, Grankin2008_spottyTTSsurvey, GullySantiago2017_youngspottystar, Feinstein2020_youngstarflares}, but is inconsistent with the 20\% covering fraction inferred from the stellar spectrum. %Regardless, the specific retrieved TLSE parameters do not impact the inferred H$_2$O abundance.

Regardless, we find consistent results for the atmospheric properties between the models with informed and uninformed stellar heterogeneity priors. The water abundance is slightly higher when using uninformed priors, at $\log_{10}\left(X_{\rm H_2O}\right) = -1.59^{+0.60}_{-0.65}$, but still within 1$\sigma$ agreement with the informed priors model. The thermal and aerosol properties are also consistent between each case. 

On the contrary, we find the inferred planet mass, $M_p$, is larger and less constrained when using uninformed priors. From this model, we find  $M_p = 11.6^{+4.2}_{-3.8}\,M_\oplus$ compared to $M_p = 8.1^{+2.5}_{-2.4}\,M_\oplus$ in the informed priors case. Since the chemical, thermal, and aerosol parameters are otherwise consistent between these two cases, this difference appears to be driven by the differing stellar heterogeneity parameters. Both retrieved values are higher than the dynamically constrained value of $4.7 \pm 0.6\,M_\oplus$ based on TTVs \citep{livingston2026_v1298tau}, but still broadly consistent with it at 1.3$\sigma$ for the informed priors result, and 1.8$\sigma$ for the uninformed priors result. Altogether, these results highlight the difficulty in deriving accurate masses for small planets around active stars. Radial velocity observations of this system have been difficult to reliably model due to significant stellar activity, and could only place a high upper limit of $M_p < 76 M_\oplus$ \citep{suarez_mascareno_2022, blunt2023_v1298tauRVs}. On the other hand, several studies have suggested that TTV analyses may underestimate planet masses due to either systematic or observational biases \citep{mills2017, leleu2023, adibekyan2024}. Retrieval-derived mass measurements, which are based on the observed atmospheric scale height, have been proposed as an alternate solution to this problem \citep{deWit2025}. However, these are still subject to challenges including degeneracies with aerosols and metallicity, and the achievable signal-to-noise for small planets. Fortunately, our inferences of the atmospheric chemistry are robust to this uncertainty on the true mass.

We do not find evidence for a high-altitude cloud deck in the atmosphere of V1298~Tau~c, with cloud deck pressures constrained to $\gtrsim10^{-2}$~bar across model considerations. However, our results prefer the inclusion of a haze scattering slope, with $\log_{10} a=7.37^{+1.56}_{-1.64}$, $\gamma=-8.80^{+2.18}_{-2.35}$ and $\phi_{\rm c+h}=0.62^{+0.12}_{-0.11}$ for the informed TLSE prior model. The aerosol parameters in the model are chosen primarily for flexibility, meaning that we are not able to infer specific haze properties such as particle composition/size from the present models. However, an aerosol slope is preferred over a slope imparted by TLSE, since any slope imparted by the TLSE will also add H$_2$O features that are not an appropriate shape to fit the observations.

\subsection{Ruling out Transit Light Source Effects} \label{subsec:retrievals_TLSE}

We obtain very weak evidence that the TLSE is impacting the transmission spectrum of V1298~Tau~c. Our TLSE model is preferred over a no-TLSE model with $\Delta \ln \mathcal{Z}=0.71, 0.51$ using the informed and uninformed priors, respectively. This suggests that the observed H$_2$O absorption features in the spectrum originate from the planet, rather than the star. Our models in which H$_2$O is removed from the atmosphere provide further insight into this finding. Models with an informed prior result in a minimal contribution from the TLSE due to the relatively small coverage fraction imposed by the prior. Models with an uninformed prior and no H$_2$O prefer larger coverage fractions and a higher spot temperature compared to the analysis in Section \ref{sec:stellarspectrum}: $f_{\rm het}=0.478^{+0.015}_{-0.029}, \, T_{\rm het}=3922^{+124}_{-123}$~K. However, even these models do not lead to recognizable H$_2$O absorption features.

\begin{figure}
    \centering
    \includegraphics[width=\linewidth]{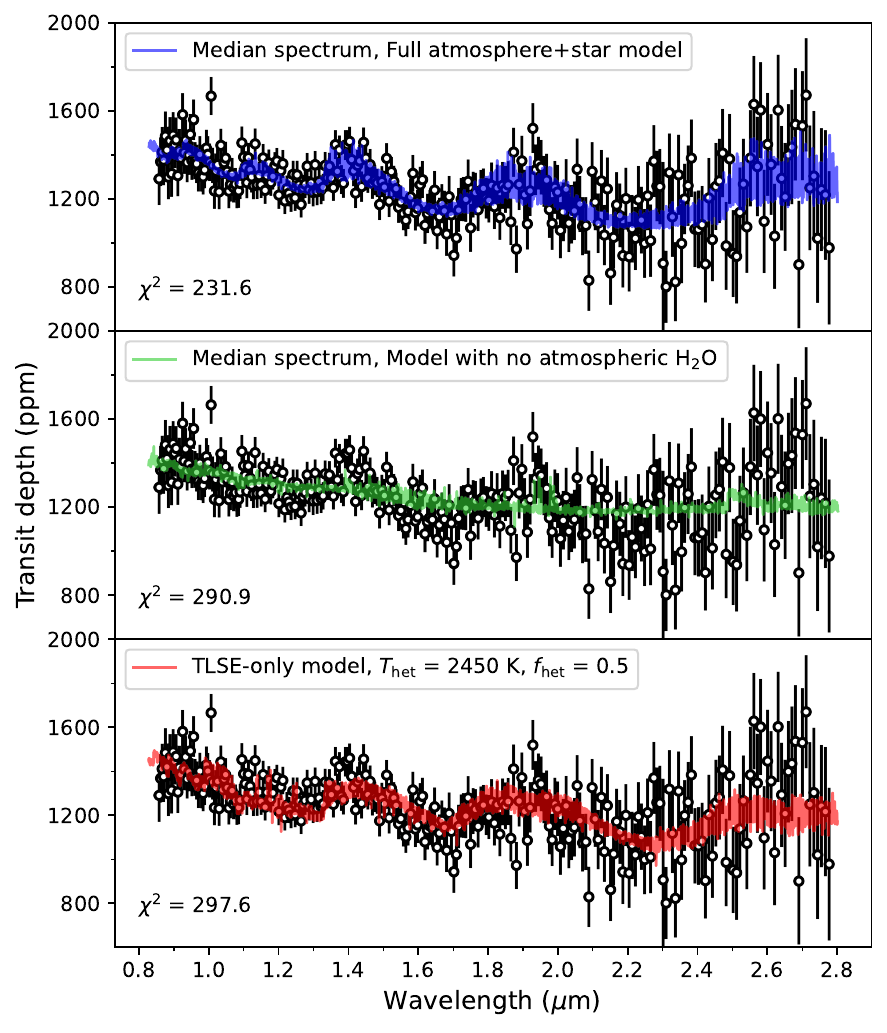}
    \caption{Exploring the limits of the contribution of stellar heterogeneities to the observed transmission spectrum. All panels show our fiducial transmission spectrum at $R$=300. 
    Top: The median retrieved model with atmospheric H$_2$O with informed stellar heterogeneity priors (same model as in Figure~\ref{fig:retrievals_spectrum}) explains the observations well.
    Middle: The median retrieved model where atmospheric H$_2$O is excluded, and we use the uninformed priors on the stellar heterogeneity parameters (Table~\ref{tab:retrievalpriors}). This TLSE-only model, even when using realisitic stellar properties, cannot explain the observations. 
    Bottom: A forward model that exhibits the maximal effect of TLSE, which requires an unphysically low starspot temperature. Even in this extreme case, the TLSE cannot explain the observations.  
    \href{https://github.com/kronos-jwst/KRONOS-I-V1298-Tau-c-NIRISS/blob/main/Create_Figure7.ipynb}{\githubicon}
    }
    \label{fig:TLSEexploration}
\end{figure}

A lack of TLSE-induced H$_2$O absorption is unsurprising given the temperature of V1298~Tau \cite[$T_\textrm{eff} = 4970 \pm 120$;][]{david19_b}. To further bolster confidence in the observed H$_2$O being planetary, we extended our models to determine what parameter space \textit{would} lead to the TLSE explaining the spectral features. We generated a grid of forward models where all wavelength-dependent atmospheric signals are removed (i.e. a constant transit depth). In turn, this means that all spectral features in this grid can only be imparted by the TLSE. Our model grid includes values of $f_{\rm het} = [0, 0.5]$ in steps of 0.05, and $T_{\rm het} = [2450, 7350]$~K in steps of 490~K. We approximate the size of the 1.4~$\mu$m feature by calculating the difference in mean transit depth from 1.35-1.45~$\mu$m and from 1.2-1.3~$\mu$m. We present examples of these TLSE grids in Figure~\ref{fig:TLSEexploration}. 

Average umbral and penumbral temperatures on the Sun range from $250-1900$~K cooler than the photosphere \citep{Solanki2003}. We find that the H$_2$O absorption feature reaches at least half the size of the feature from the overall best-fit model only for models with unphysical spot properties: $T_{\rm het}=2450$~K ($f_{\rm het} \geq 0.35$) and 2940~K ($f_{\rm het} \geq 0.45$). These temperatures are low enough that some, but not all, H$_2$O would be thermally dissociated. An example of the former is shown in the bottom panel of Figure~\ref{fig:TLSEexploration}. While the spot coverage fractions may be feasible \citep{GullySantiago2017_youngspottystar}, these spots would be $\Delta T$=1960--2450\,K cooler than the stellar photosphere, which is more extreme than has been seen of spot umbrae on the Sun \citep[$\Delta$T=972–1872; ][]{Solanki2003}.

Additionally, models with H$_2$O features due to the TLSE provide a worse fit to the data overall when compared to models with H$_2$O in the planetary atmosphere (top panel of Figure~\ref{fig:TLSEexploration}), as well as TLSE-only models that do not contain stellar water features (middle panel of Figure~\ref{fig:TLSEexploration}). This is due to the different shape of water features resulting from the TLSE, which originate from starspots at much higher temperatures ($T>2000$~K) compared to the planetary atmosphere. Therefore, we conclude that the TLSE assuming stellar properties expected for V1298~Tau cannot explain the observed transmission spectrum, and thus the observed H$_2$O features in the transmission spectrum of V1298~Tau~c must originate from its atmosphere.

\subsection{Comparison to Previous HST Measurements} \label{subsec:atmomodeling_discussion_HSTcomparison}

\begin{figure}
    \centering
    \includegraphics[width=\linewidth]{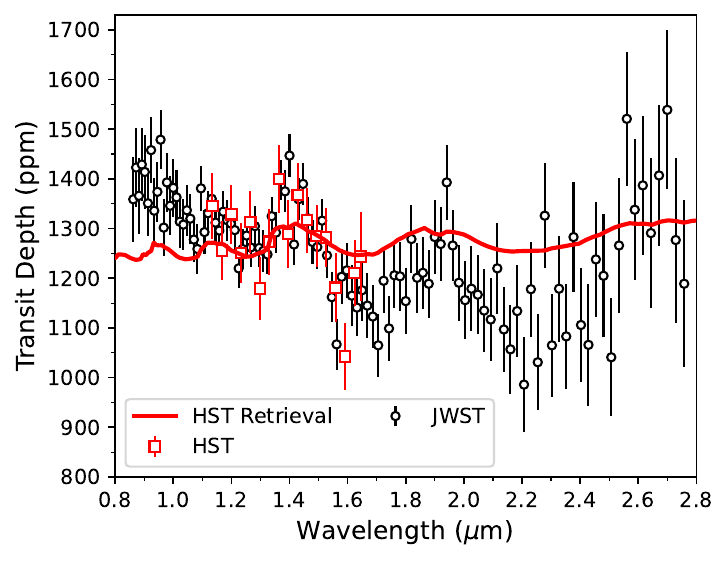}
    \caption{Comparison between our JWST NIRISS/SOSS transmission spectrum (black) and the HST WFC3/G141 transmission spectrum and median retrieved model (red) presented by \protect\cite{barat2024_v1298taubc}. We find the absolute transit depth is consistent between observations. The extended wavelength coverage and increased precision of NIRISS/SOSS allows for a more confident inference of H$_2$O in V1298~Tau~c.
    \href{https://github.com/kronos-jwst/KRONOS-I-V1298-Tau-c-NIRISS/blob/main/Create_Figure8.ipynb}{\githubicon}
    }
    \label{fig:NIRISSvsHST}
\end{figure}

\cite{barat2024_v1298taubc} previously measured the $\sim$1.1--1.6\,$\mu$m transmission spectrum of V1298~Tau~c using HST WFC3/G141 on UT 2021 October 18. We compare the observed transmission spectra between HST and JWST Figure~\ref{fig:NIRISSvsHST}, along with the best-fit retrieved model of \cite{barat2024_v1298taubc}. The absolute transit depth and relative depths around the 1.4$\mu$m H$_2$O feature are consistent between visits and facilities. 
%Although we cannot say for certain, the consistency in absolute transit depth between these visits suggests the properties of surface heterogeneities on V1298~Tau may have been very similar at these distinct epochs. 

As discussed by \cite{barat2024_v1298taubc}, the HST spectrum exhibits relatively higher scatter and transit depth uncertainties, most likely due to a very limited pre-transit baseline. Combined with its limited wavelength coverage, this led \cite{barat2024_bHST} to conclude V1298~Tau~c has a featureless spectrum. In fact, \cite{barat2024_v1298taubc} found only a moderate ($\sim$2.5$\sigma$) preference for a model with atmospheric water absorption versus a flat line model using a $\chi^2$ test. Assuming a mass of 10~$M_\oplus$, consistent with the mass inferred from our models, \cite{barat2024_v1298taubc} inferred a near-solar metallicity of $\log\left(Z\right) = -0.02^{+1.3}_{-1.2}$ and hypothesized significant atmospheric haze formation to explain the small feature size. The increased precision and wavelength coverage of NIRISS/SOSS allows for a more confident inference of atmospheric H$_2$O and refinement of the atmospheric metallicity to a super-solar value.

\subsection{No Evidence for Limb Asymmetry} \label{subsec:atmomodeling_discussion_LAdiscussion}

%\subsection{Exploration of Limb Asymmetry}
%\label{subsec:lcfitting_catwoman}

Up until this point, our transit light curve fitting has assumed V1298~Tau~c has uniform limbs, implying that the atmospheric properties at the evening and morning terminators are similar or equivalent. Limb asymmetry has recently been detected on a handful of planets with JWST \citep[][c.f., \cite{radica_super-solar_2026}]{espinoza2024_wasp39b, murphy2024_wasp107b, ahrer2025_wasp94b, mukherjee2025_wasp94b, wang2026_wasp96b}. The prevalence and origins of such asymmetry are not fully understood \citep{fu2025}. 

\begin{figure}
    \centering
    \includegraphics[width=\linewidth]{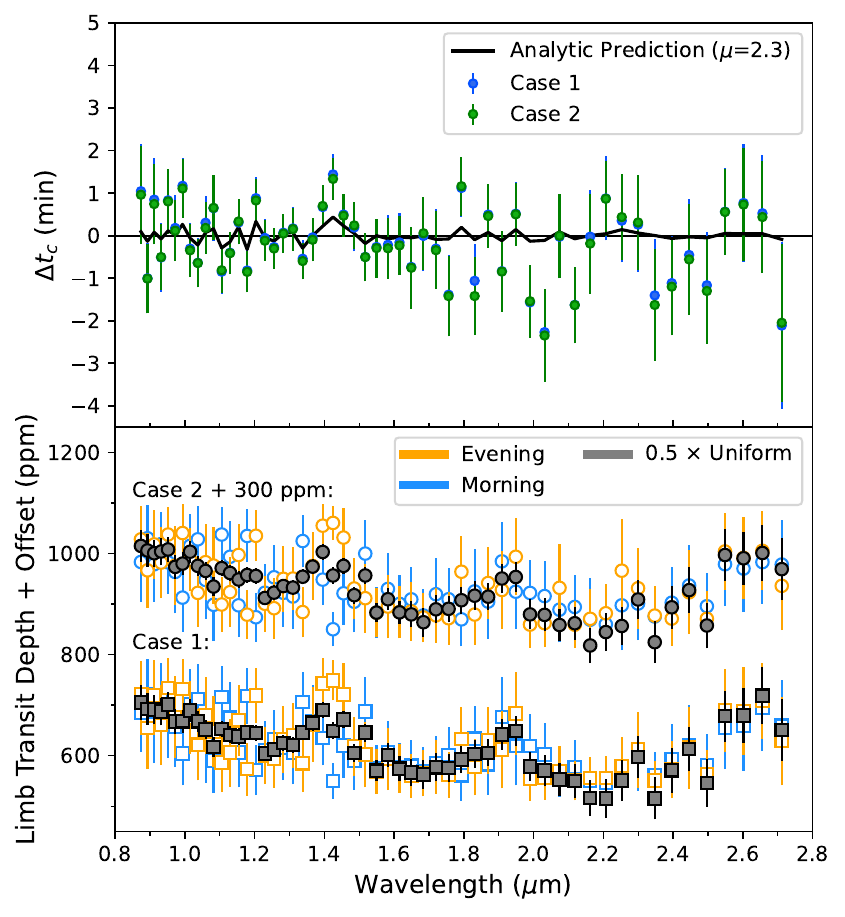}
    \caption{Results of tests for evening-morning limb asymmetry on V1298~Tau~c. 
    Top: The timing bias spectra from our uniform-limb spectroscopic fits in Cases 1 (blue) and 2 (green), compared to the analytic model (black line) of \protect\cite{Murphy2024_tcbias}. 
    Bottom: The evening- (orange) and morning-limb (blue) spectra compared to their corresponding uniform-limb spectra (gray). The Case 2 results are all offset by 300~ppm for visual clarity, but the evening and morning limb spectra are not offset relative to one another. We see no significant evidence for limb asymmetry on V1298~Tau~c.
    \href{https://github.com/kronos-jwst/KRONOS-I-V1298-Tau-c-NIRISS/blob/main/Create_Figure9.ipynb}{\githubicon}
    }
    \label{fig:limbasymmetry}
\end{figure}

To investigate limb asymmetry on V1298~Tau~c, we perform two parallel tests. First, we freely fit for $t_c$ as a function of wavelength in the spectroscopic light-curve fits assuming uniform limbs. Second, we repeated the spectroscopic fits for Cases 1 and 2 using \texttt{catwoman} \citep{catwoman, espinoza2021_la} as the transit model. \texttt{catwoman} splits the occulting disk of the planet into two semi-circles with independent radii, as opposed to \texttt{batman} which assumes the disk is a uniform circle. We conducted these fits at $R=50$ to minimize light curve noise. The overall methodology was identical, except that  $t_{\mathrm{c}}$ was fixed to its broadband value, and we included two $R_{\mathrm{p}}/R_\star$ parameters (one for each limb) instead of one. Also, we assumed that the planetary obliquity $\phi$ is 90$^{\circ}$, meaning the planet's rotation and orbital axes are aligned. We did not test for limb asymmetry using our Case 3 and 4 fitting routines as \texttt{catwoman} is not implemented in \texttt{fleck} and overfitting with the GP could provide ambiguous results. The results of these tests are presented in Figure~\ref{fig:limbasymmetry}.

We find no strong evidence for evening--morning limb asymmetry on V1298~Tau~c. The top panel of Figure~\ref{fig:limbasymmetry} shows the timing bias $\Delta t_c$ spectrum from the masked (Case 1) and Gaussian fits (Case 2), obtained from the same uniform-limb fits that derived the transmission spectra presented in Figure~\ref{fig:transmissionspectra}. These timing biases are the difference between the best-fit $t_c$ at each wavelength and the broadband-derived value. The bottom panel of Figure~\ref{fig:limbasymmetry} shows the evening- and morning-limb transmission spectra obtained from our \texttt{catwoman} fits, compared to their respective uniform-limb spectra. 

The $\Delta t_c$ spectrum is featureless and generally consistent with zero at all wavelengths, except one possible feature near 1.4~$\mu$m. Similarly, the evening- and morning-limb spectra exhibit no clear asymmetry except in a few channels near 1.4~$\mu$m. Although these two possible hints at asymmetry do align, they are likely noise-driven. The 1.4~$\mu$m feature is a strong water absorption feature, but there are several others in this bandpass (e.g., at 1.8~$\mu$m) that do not exhibit the same asymmetry, inconsistent with a significant thermal or abundance gradient between terminators. The identical slopes shortward of this feature similarly argue against a significant difference in aerosol properties as well, with the caveat of the TLSE contribution to this slope. Furthermore, the timing biases are inconsistent in magnitude with the limb-limb differences in transit depth. The black line in the top panel of Figure~\ref{fig:limbasymmetry} shows an analytic prediction for the timing bias spectrum based on the depth differences measured in Case 2, assuming a mean molecular weight of $\mu$=2.3, following the method of \cite{Murphy2024_tcbias}. For true atmospheric signals, this prediction should match the measured values since the timing bias is geometrically caused by the difference in the projected radius of each limb. However, the measured values near 1.4~$\mu$m are larger by a factor of 2. 

V1298~Tau~c having a homogeneous atmosphere would be generally consistent with predictions from cloud-free general circulation models \citep{kataria2016, roth2024}. However, no matter which mass estimate we use, V1298~Tau~c lies below the ``asymmetry horizon'' in temperature--gravity space inferred by \cite{fu2025}. This places V1298~Tau~c in the same regime where other planets (WASP-39~b, \citeauthor{espinoza2024_wasp39b} \citeyear{espinoza2024_wasp39b}; WASP-94~Ab, \citeauthor{mukherjee2025_wasp94b} \citeyear{mukherjee2025_wasp94b}; WASP-17~b, \citeauthor{fu2025} \citeyear{fu2025}) do exhibit significant limb asymmetry in the form of both thermal and aerosol gradients. There are two possible systematic biases that could be masking the signal of limb asymmetry in our data. First, there are starspot crossings throughout nearly the entire transit, which are known to bias the measured limb depths from a \texttt{catwoman} fit if improperly accounted for \citep{Murphy2025_wasp107b}. Since we test multiple ways of modeling the SCEs which yield consistent results though, this bias should be minimal. Second, there are significant transit timing variations in the V1298~Tau system, which can exceed 2 hours for V1298~Tau~c \citep{livingston2026_v1298tau}. Since V1298~Tau~c's time of conjunction changes from transit to transit relative to a linear ephemeris, there is no single value that can be measured precisely from multi-epoch multi-wavelength observations. As a result, $t_c$ can only be measured from the same data used to investigate limb asymmetry, and is therefore subject to bias from any underlying limb asymmetry that, in return, masks the asymmetry \citep{Powell2019, Murphy2024_tcbias}. If this bias is present, it would primarily cause a vertical offset (or lack thereof) between the derived limb spectra. A thorough investigation of which bias is impacting our data the most is reserved for future work.

\section{Implications for Planetary Evolution} \label{sec:formation}

From our derived NIRISS/SOSS transmission spectrum, we can begin to extrapolate trends in atmospheric composition as a function of mass, equilibrium temperature, and age.

\subsection{Mass--Metallicity Evolving in Time} \label{subsec:formation_massmetallicity}

A relation between a planet's mass and atmospheric metallicity has been sought after as a link between planet formation and present-day observables \citep{thorngren2016, welbanks2019, Sun2024, swain2024}. However, the mass and composition of exoplanet atmospheres can change significantly from their initial state due to various evolutionary processes early in their lifetime. Therefore, trends inferred from observations of mature planets likely do not reflect the primordial mass--metallicity relation. Young exoplanets, on the other hand, offer a unique opportunity to recover the primordial mass--metallicity relation. 

%The atmospheric metallicity of V1298~Tau~c we estimate from our free retrievals is based on its H$_2$O abundance, which yields an O/H metallicity value of approximately 15$\times$ Solar. This is generally consistent with the H$_2$O-based mass-metallicity trends found by \cite{welbanks2019} and \cite{Sun2024}, which predict $\sim$11$\times$ Solar and 6$\times$ Solar for V1298~Tau~c, respectively. Our upcoming JWST/NIRSpec observation will improve on this by providing access to carbon- and sulfur-bearing species, which will enable a more robust measurement of the bulk atmospheric metallicity, and comparison of individual elemental enrichments. 

\begin{figure}
    \centering
    \includegraphics[width=1\linewidth]{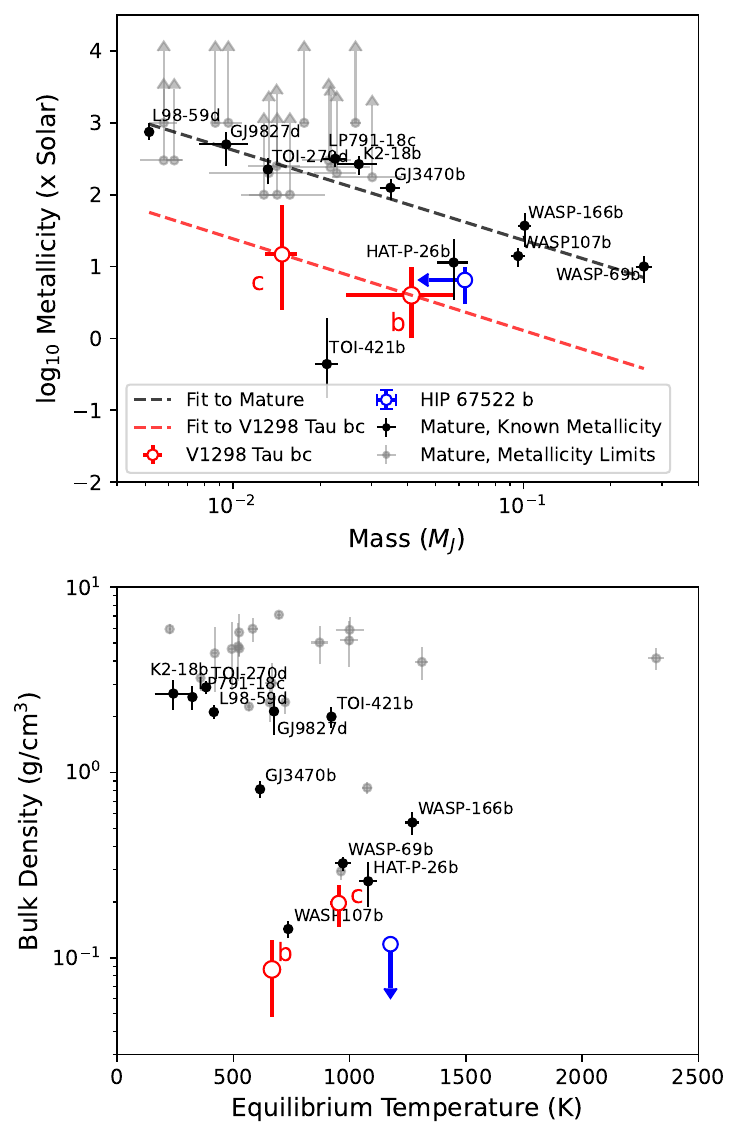}
    \caption{Placing V1298~Tau~c into context with other young and mature exoplanets. Top: atmospheric metallicity versus planet mass. The masses for V1298~Tau~bc are from \cite{livingston2026_v1298tau}. Bottom: bulk density versus equilibrium temperature. We present systems with constrained ages in color and assume systems without ages are mature ($>1$~Gyr). Planets with lower limits on atmospheric metallicity are plotted in gray. Based on our observations, we speculate that the mass--metallicity relation evolves with planetary age due to early-time atmospheric evolution. Additional observations are required to more robustly explore this trend.
    \href{https://github.com/kronos-jwst/KRONOS-I-V1298-Tau-c-NIRISS/blob/main/Create_Figure10.ipynb}{\githubicon}
    }
    \label{fig:massmetallicity}
\end{figure}

We place the inferred metallicity of V1298~Tau~c and b into context with their mature counterparts, which we define as planets with $>1$~Gyr or unconstrained ages but with similar masses as our targets. V1298~Tau~b is the only other planet in this system with a published metallicity estimate \citep{barat2025_v1298taub}. We also highlight HIP~67522~b, another young ($\sim$17~Myr) super-Neptune \citep{thao2024}. We present a synthesis of mass--metallicity information in Figure~\ref{fig:massmetallicity}.  We focus on measurements based on JWST observations (see Appendix~\ref{apx:metallicities} for references), and it is important to note that these measurements are derived using a variety of methodologies and are based on different combinations of elements. Nevertheless, we see a new clear trend in metallicity for planets of the same mass, but very different ages.  All three young planets have relatively low metallicities on the order of 1-10$\times$ the Solar value. By contrast, the mature exoplanets have significantly higher metallicities. TOI-421~b is a notable exception, though their data does allow for higher metallicity solutions of $\sim$100$\times$ solar, which would be consistent with the observed trend \citep{davenport2025}. Due to the challenge of observing atmospheres around lower mass super-Earths, many of their measurements are only lower limits (gray points). Regardless, we can fit the constrained metallicities $Z$ of the mature planets as a function of mass with a line of the form
\begin{equation}
    \log_{10}\left(Z\right) = \left(-1.25 \pm 0.21\right) \log_{10}\left(\frac{M_p}{M_J}\right) + \left(0.11 \pm 0.33\right). \label{eqn:matureZfit}
\end{equation}
Here, $M_p$ is in units of Jupiter masses, $M_J$. This fit excludes TOI-421~b due to its anomalously low metallicity. Similarly, we can fit the metallicities within the V1298~Tau system \citep[][and this work]{barat2025_v1298taub} with a line of form
\begin{equation}
    \log_{10}\left(Z\right) = -1.28 \log_{10}\left(\frac{M_p}{M_J}\right) - 1.17. \label{eqn:matureZfit}
\end{equation}
The latter has no coefficient uncertainties since it is simply a fit between two points. We exclude HIP~67522~b in this fit due to its uncertain mass. These two trends are displayed in Figure~\ref{fig:massmetallicity}. The mass--metallicity slopes for the mature planets and young V1298~Tau planets are near-identical, with only a vertical (i.e., metallicity) offset between them. 

This tentative offset between the young and mature mass--metallicity trends may directly reflect the outcome of atmospheric evolution. Hydrogen and helium, which make up the bulk of the primordial atmosphere, are preferentially lost through boil-off and photoevaporation. This will naturally lead to an increase in atmospheric metallicity over time \citep{Malsky2020, louca2025, valatsou2026}, even if there is a metallicity gradient within the system depending on the planet's formation location \citep{ormel2021,bloot2023}. We also see this scenario in terms of the planet bulk densities (bottom panel of Figure~\ref{fig:massmetallicity}), where the young planets have relatively extreme low densities because they still have extended hydrogen and helium envelopes in addition to significant internal heat. Thus, we expect the mass--metallicity relation will be a gradient toward higher metallicity with increasing planetary age. This may also imply that the diverse set of generally high metallicities seen on mature super-Earths and sub-Neptunes are not primordial, but instead are the product of billions of years of atmospheric evolution.

%Atmospheric metallicity could increase through preferential loss of lighter elements like H and He by an order of magnitude \citep{Malsky2020,louca2025,valatsou2026}. Alternatively, young sub-Neptunes could be born with metallicity gradients \citep{ormel2021,bloot2023}, similar to solar system giants \citep{wahl2017,vazan2020}. If such an envelope is subjected to atmospheric mass loss, it would remove the metal-rich outer layers and reveal the high metallicity inner layers with age, resulting in atmospheric metallicity evolution. 
Recent modelling of atmosphere-interior interactions have also shown that significant amount of volatiles could be dissolved in global magma oceans. This may influence the initial atmospheric ratio of H/He to metals, and lead to further evolution as volatiles outgas over time \citep{kite2020,werlen2025,steinmeyer2026, kimura2026}. These interactions likely also have an effect on how the atmospheric metallicity evolves with age, particularly in the first $\sim$100~Myr \citep{kimura2026}. We will test these hypotheses with new JWST observations of V1298~Tau~d, TOI-451~cd (120~Myr) and TOI-2076~bcd (250~Myr) as part of the KRONOS program. Additionally, forthcoming observations of younger and similarly aged systems will be able to better constrain the observed trend presented here  \citep[e.g. JWST GO 5311, 6670, 8597;][]{feinstein24_jwst, louca24, feinstein25_jwst}.

\subsection{Comparison to Mature Super-Earths} \label{subsec:formation_SEcomparisons}

V1298~Tau~c's extremely low density makes it an ideal laboratory for understanding the conditions under which super-Earths ($1.2 R_\oplus < R_p \leq 1.8 R_\oplus$) form. Using the refined mass from \cite{livingston2026_v1298tau} and measured X-ray flux of V1298~Tau from \cite{poppenhaeger21}, V1298~Tau~c is expected to evolve to $R_p \lesssim 2R_\oplus$. The only mature super-Earth orbiting an FGK star to have received atmospheric characterization thus far is TOI-561~b (R$_p$ = 1.4\,R$_\oplus$, T$_{\mathrm{eq}}$ = 2300\,K, R$_\star$ = 0.843\,R$_\odot$). \cite{teske2025_TOI561b} presented the 3--5\,$\mu$m emission spectrum of TOI-561~b via JWST/NIRSpec, which provides evidence that TOI-561~b retains a relatively thick, volatile-dominated atmosphere despite lying well above the predicted ``cosmic shoreline'' \citep{zahnle17}. However, it is not clear whether this atmosphere is primordial or secondary. V1298~Tau~c and TOI-561~b represent different life stages of the same population of planets. By observing comparable-mass planets around young, intermediate, and mature G stars, we can better understand the affects of photoevaporation, photochemistry, and secondary atmospheres in the vast population of super-Earths.

By contrast, transmission observations of mature super-Earths orbiting M stars have predominantly revealed featureless or tentative detections of molecules in their atmospheres. This includes GJ~486~b \citep{Moran2023_GJ486b}, L~98-59~c \citep{scarsdale2024_L9859c}, L~98-59~d \citep{gressier2024_l9859d}, LTT~3780~b \citep{allen2025_LTT3780b}, LTT~1445~Ab \citep{diamondlowe2023_LTT1445Ab, wachiraphan2025_LTT1445Ab}, and TOI-1685~b \citep{luque2025_TOI1685b}. In some cases, the data favors a bare rock scenario \citep[e.g.,][]{allen2025_LTT3780b, wachiraphan2025_LTT1445Ab}, while other cases are more ambiguous and could still allow a thin atmosphere \citep{Moran2023_GJ486b, scarsdale2024_L9859c, luque2025_TOI1685b}. These planets make up the bulk of upper limits shown in the top panel of Figure~\ref{fig:massmetallicity}. Since the XUV environment close in to M stars is relatively more intense than for FGK systems \citep[e.g.,][]{shkolnik2014, france2016, loyd2018, peacock2020, loyd2021}, these planets may have been stripped more than V1298~Tau~c inevitably will. However, without any confirmed atmospheres, no definitive conclusions can be drawn. Future transmission observations of young and intermediate aged super-Earths around M dwarfs are necessary to draw trends in the evolution of planets around stars of different spectral types.

%===========================================================%
\section{Summary} \label{sec:summary}

We present the $\sim$0.85--2.83\,$\mu$m transmission spectrum of V1298~Tau~c, a $\sim$23~Myr super-Earth or sub-Neptune progenitor, derived from one transit observation with JWST NIRISS/SOSS as part of the KRONOS program (JWST GO 5959). We summarize our key findings as the following.

\begin{enumerate}
    \item Despite the presence of at least six starspot crossings during our transit, we derive highly consistent transmission spectra from four independent methods for modeling the transit of V1298~Tau~c: 1) masking the starspot crossings, 2) modeling them as Gaussian profiles, 3) modeling them using \texttt{fleck}, and 4) modeling them using a Gaussian Process model. These results prove that characterizing exoplanetary atmospheres is possible in the face of significant stellar surface heterogeneities.
    
    \item We find evidence for H$_2$O in V1298~Tau~c's atmosphere at high statistical significance ($\Delta \ln \mathcal{Z} \geq 27.25$), and infer an abundance of $\log_{10}\left(X_{H_2O}\right)$ = -1.83$^{+0.68}_{-0.77}$. This corresponds to an atmospheric metallicity of $\log_{10}\left(\mathrm{O/H}\right) = 1.17^{+0.68}_{-0.77}$, or $14.8^{+56.0}_{-12.28}$ times the Solar value. Upcoming observations with JWST/NIRSpec G395H ($\sim$3--5~$\mu$m) will refine this value, and additionally constrain the ratios of carbon and sulfur. 
    
    \item We fit the out-of-transit stellar spectrum to derive the surface heterogeneity properties of V1298~Tau. Beyond a quiet $\sim$4900\,K photosphere, we find a cool component (e.g., spots) of $\sim$3400\,K that covers up to $\sim$20\% of the stellar surface, and a hot component (e.g., faculae and plages) of $\sim$5600\,K that covers up to $\sim$30\% of the surface. While the coverage fractions differ, the temperatures are consistent with modern day solar values.
    
    \item We consider atmospheric retrieval models with and without informed priors on the stellar surface heterogeneities from our stellar spectrum fit. The retrieved atmospheric parameters are highly consistent. In particular, the H$_2$O abundances are consistent within 0.5$\sigma$, suggesting that whether free or informed priors are used, the models still prefer the H$_2$O originating in the planet's atmosphere.
    
    \item We find no evidence for significant evening-morning limb asymmetry on V1298~Tau~c, though the signal of limb asymmetry in our data may be masked by the numerous starspot crossings during transit and/or transit timing variations in this system.
    
    \item V1298~Tau~c joins a growing list of young planets with relatively low metallicities compared to mature planets of similar masses. We find the mass--metallicity trend for young planets may show a vertical offset from mature planets, but the slope may remain the same. 
    However, this tentative trend is based on only three measurements from two young planetary systems, and additional observations of other young systems are necessary. Another caveat is that this comparison collects measurements from varied methodologies and are based on different elemental ratios, which motivate a more thorough analysis. Nevertheless, these results suggest that the planetary mass--metallicity trend may change with age due to atmospheric evolutionary processes, which will be illuminated by future KRONOS program.

\end{enumerate}

This work highlights the interdisciplinary insights which can be achieved by observing the transits of planets orbiting young active stars. We demonstrate that robust atmospheric characterization of young exoplanets is achievable with JWST, even in the presence of significant starspot crossing events. In this case, we are aided in breaking the degeneracy between stellar and planetary signals by the distinct, well-separated temperatures of the stellar components and the absence of water features in the starspot spectra due to the warmer host star $T_{\mathrm{phot}}$, particularly compared to M dwarfs which may have significant water in their starspots.

This separation will be substantially more challenging for cooler stars---low mass K and M dwarfs--- spot--photosphere temperature contrasts are smaller \citep{berdyugina2005} and the spectral features induced by heterogeneities can be more similar to features expected in the planetary atmosphere. This has already been an issue for observations of multiple systems \citep[e.g., GJ~486, GJ~1132~b, and TOI-5205~b;][]{Moran2023_GJ486b, may2023_gj1132, canas2026_TOI5205b}.
In those cases, simultaneous and follow-up optical and near-infrared spectrophotometric observations from the ground and/or new missions like Pandora \citep{Barclay2025, Rackham2026, Rotman2026} will be critical to enable reliable atmospheric characterization across the full range of host star types. 

The observations and models that are presented as key results in this manuscript are publicly available online at \url{https://github.com/kronos-jwst/KRONOS-I-V1298-Tau-c-NIRISS}. All the data used in this paper can be found in MAST: \dataset[10.17909/10.17909/vjpc-v080]{http://dx.doi.org/10.17909/vjpc-v080}

%% Please use the acknowledgment and contribution environments. This will 
%% be anonomyized when the "anonymous" style option is used. 
\begin{acknowledgments}

The authors would like to thank the co-Is who contributed to the KRONOS proposal: Eva-Maria Ahrer, Lili Alderson, Jonathan Brande, Jean-Mich\'el D\'esert, N\'estor Espinoza, Peter Gao, Giannina Guzman Caloca, Garrett Levine, Andrew Mann,  James Owen, Keighley Rockcliffe, Leslie Rogers, Sara Seager, Alexander Shapiro, Pa Chia Thao, and Shreyas Vissapragada. The authors would also like to thank Richard Booth for reviewing an earlier version of this manuscript, which helped improve the presentation of our results, as well as our anonymous referee, whose thorough review and helpful suggestions greatly improved our analysis. 

This work is based on observations made with the NASA/ESA/CSA JWST. The data were obtained from the Mikulski Archive for Space Telescopes at the Space Telescope Science Institute, which is operated by the Association of Universities for Research in Astronomy, Inc., under NASA contract NAS 5-03127 for JWST. The JWST data presented in this article can be accessed via  \dataset[doi: 10.17909/vjpc-v080]{https://doi.org/10.17909/vjpc-v080}.
These observations are associated with the program JWST GO 5959. Support for program JWST GO 5959 was provided by NASA through a grant from the Space Telescope Science Institute.
M.C.N. and S.M. thank the Heising-Simons foundation for their support through the 51 Pegasi b Fellowship. This work made use of High-Performance Computing Facilities at Arizona State University \citep{HPC_ASU23}.

This material is based upon work supported by the National Aeronautics and Space Administration under Agreement No.\ 80NSSC21K0593 for the program ``Alien Earths''. The results reported herein benefited from collaborations and/or information exchange within NASA’s Nexus for Exoplanet System Science (NExSS) research coordination network sponsored by NASA’s Science Mission Directorate. This material is based upon work supported by the European Research Council (ERC) Synergy Grant under the European Union’s Horizon 2020 research and innovation program (grant No.\ 101118581---project REVEAL). This work is partly supported by JSPS KAKENHI Grant Numbers JP24H00017 and JP25K17450. R.L. is funded by the European Union (ERC, THIRSTEE, 101164189) and acknowledges financial support from the Severo Ochoa grant CEX2021-001131-S funded by MCIN/AEI/10.13039/501100011033. Part of this work was carried out at the Jet Propulsion Laboratory, California.

\end{acknowledgments}

\begin{contribution}
M.M.M. led the data analysis and wrote this article. M.C.N. led the modeling interpretation of the data and contributed to writing the article. A.D.F. and L.W. led the original proposal for these data, led the data collection, and contributed to the interpretation of our results and writing this article. G.M.D., S.B., B.V.R., and D.Z.S. aided in the data analysis, interpretation, and contributed to writing this article. M.R. aided in the data reduction. All other authors contributed to the original proposal for these data, contributed to its interpretation, and aided in preparing this manuscript.
\end{contribution}

\facilities{JWST(NIRISS)}

\software{
\texttt{numpy} \citep{numpy},
\texttt{matplotlib} \citep{matplotlib},
\texttt{scipy} \citep{scipy},
\texttt{exotedrf} \citep{exotedrf},
\texttt{batman} \citep{batman},
\texttt{catwoman} \citep{catwoman},
\texttt{speclib} \citep{Rackham2023, Rackham2024},
\texttt{UltraNest} \citep{Buchner2014_pymultinest},
\texttt{PyMultiNest} \citep{Buchner2014_pymultinest}
}

% %%%%%%%%%%%%%%%%%%%%%%%%%%%%%%%%%%%%%%%%%%%%%%%%%%
% \section*{Data Availability}
% The JWST data used in this article is publicly accessible from the Mikulski Archive for Space Telescopes, and can be found at \textcolor{red}{[MAST DOI created upon final submission].} We have also tabulated our transmission spectra, stellar spectrum, and starspot contrast measurements which can be downloaded from \textcolor{red}{[Zenodo repository created upon final submission].}

%%%%%%%%%%%%%%%%% APPENDICES %%%%%%%%%%%%%%%%%%%%%
\clearpage
\appendix

\section{Comparison of Limb Darkening Profiles} \label{apx:LDprofiles}

\begin{figure*}
    \centering
    \includegraphics[width=\linewidth]{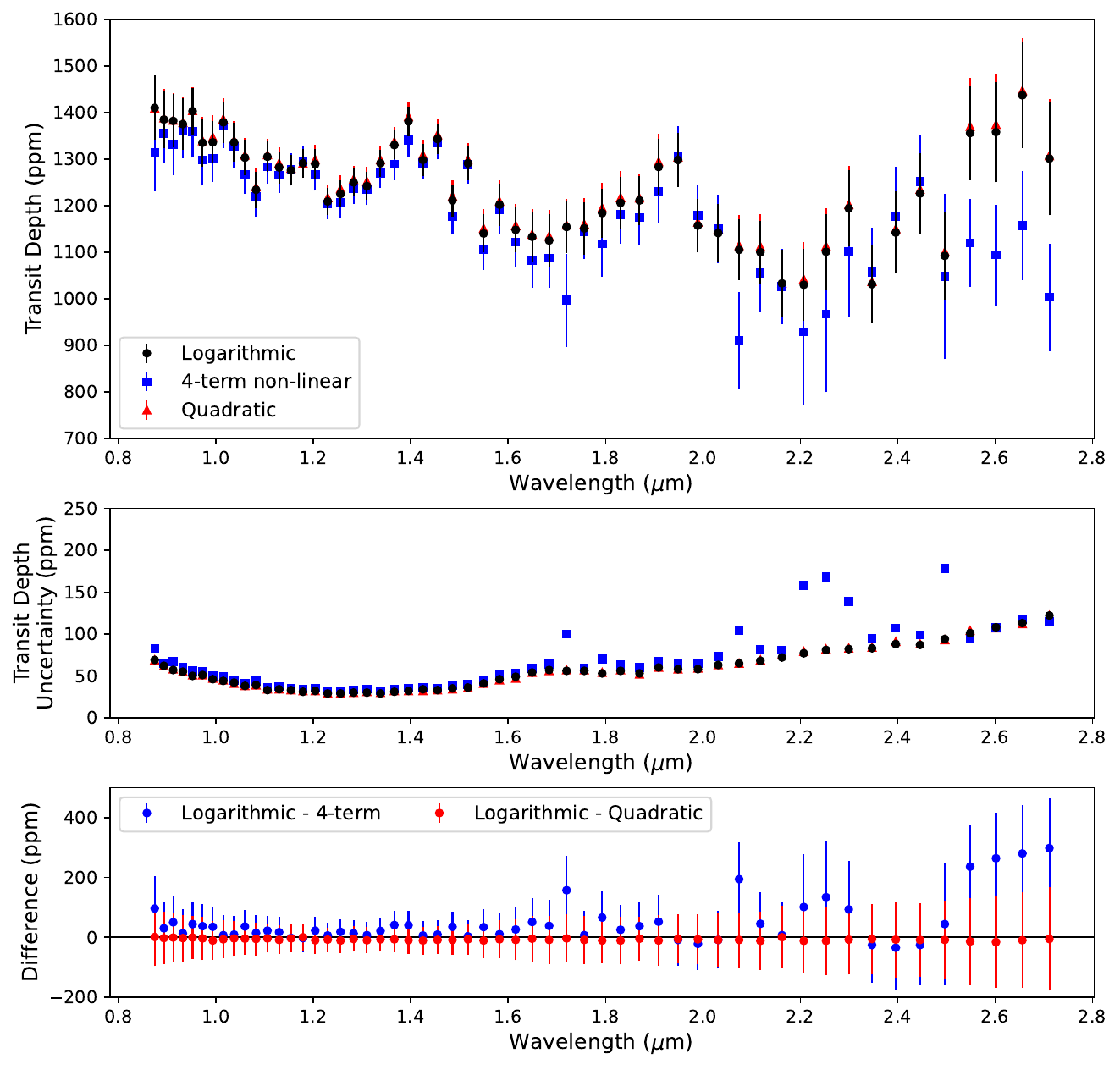}
    \caption{Top: The transmission spectrum of V1298~Tau~c, at a resolution R=50, derived assuming three different limb darkening profiles when fitting the spectroscopic light curves, and enforcing only uniform priors on the coefficients. In each case, we mask all starspot crossing features seen in the light curve. Middle: The fitted transit depth uncertainties from each case. Bottom: The difference in transit depths between the three cases. The error bars represent the mutual propagated uncertainty on the difference.
    \href{https://github.com/kronos-jwst/KRONOS-I-V1298-Tau-c-NIRISS/blob/main/Create_Figure11.ipynb}{\githubicon}}
    \label{fig:spectrumcomp_LDprofiles}
\end{figure*}

As mentioned in Section~\ref{subsec:lcfitting_limbdarkening}, we tested three different limb darkening profiles. Table~\ref{tab:bics} compares the goodness-of-fit metrics between each test, including testing different out-of-transit baseline models. A logarithmic profile was preferred based on fits to the broadband light curve, but we wanted to ensure that this choice did not have a meaningful impact on our ultimate results. We therefore derived the transmission spectrum assuming each profile in our light curve fits, following Case 1 (Section~\ref{subsec:lcfitting_case1mask}) for simplicity, shown in Figure~\ref{fig:spectrumcomp_LDprofiles}. The logarithmic results, used in our primary analysis, and the quadratic results, typically used in the literature, are nearly identical. The spectrum derived using a 4-term profile is most different from the other two, though the difference is primarily a vertical offset. Particularly toward the longer wavelengths where the data quality is relatively poorer, differences between individual channels is consistent with scatter and comparable to the individual transit depth uncertainties. 

\begin{table*}[]
    \centering
    \caption{Goodness-of-fit metrics for various baseline models and limb darkening treatments, fit to the broadband light curve with SCEs masked.}
    \begin{tabular}{c|c|c|c|c}
    Baseline & LD Prior & $\ln\left(L\right)$ & BIC & $\Delta$$|$BIC$|$ \\ \hline 
   \multicolumn{5}{c}{ \textit{Quadratic Limb Darkening}} \\ \hline 
    Quadratic & Uniform & 12194.81 & -24315.6 & 39.4 \\
    Quadratic & Gaussian & 12194.66 & -24315.3 & 39.7 \\
    Cubic & Uniform & 12218.19 & -24355.0 & 0.0 \\
    Cubic & Gaussian & 12217.55 & -24353.7 & 1.3 \\
    4-term & Uniform & 12219.40 & -24350.0 & 5.0 \\
    4-term & Gaussian & 12219.72 & -24350.6 & 4.3 \\
    5-term & Uniform & 12220.11 & -24344.0 & 11.0 \\
    5-term & Gaussian & 12220.29 & -24344.4 & 10.6 \\
    \hline \multicolumn{5}{c}{ \textit{Logarithmic Limb Darkening}} \\ \hline 
    Quadratic & Uniform & 12196.55 & -24319.1	& 37.1 \\
    Quadratic & Gaussian & 12195.17 & -24316.3 & 39.9 \\
    Cubic & Uniform & 12218.82 & -24356.2	& 0.0 \\ 
    Cubic & Gaussian & 12218.29 & -24355.2 & 1.1 \\
    4-term & Uniform & 12220.26 & -24351.7 & 4.5 \\
    4-term & Gaussian & 12219.64 & -24350.5 & 5.8 \\
    5-term & Uniform & 12220.58 & -24345.0 & 11.3 \\
    5-term & Gaussian & 12219.96 & -24343.7 & 12.5 \\
\hline \multicolumn{5}{c}{ \textit{4-term Non-linear Limb Darkening}} \\ \hline 
    Quadratic & Uniform & 12190.76 & -24292.7 & 47.8 \\
    Quadratic & Gaussian & 12195.13 & -24301.4 & 39.1 \\
    Cubic & Uniform & 12216.93 & -24337.6 & 2.9 \\
    Cubic & Gaussian & 12218.36 & -24340.5 & 0.0 \\
    4-term & Uniform & 12194.22 & -24284.8 & 55.7 \\
    4-term & Gaussian & 12220.20 & -24336.8	& 3.7 \\
    5-term & Uniform & 12219.80 & -24328.6 & 11.9 \\
    5-term & Gaussian & 12220.76 & -24330.5	& 10.0 \\
    \end{tabular}
    \tablecomments{$\ln\left(L\right)$ refers to the likelihood computed using the median of each fitting parameter's posterior distribution. The $\Delta$$|$BIC$|$ values are calculated within each family of limb darkening laws.}
    \label{tab:bics}
\end{table*}

\begin{figure*}
    \centering
    \includegraphics[width=\linewidth]{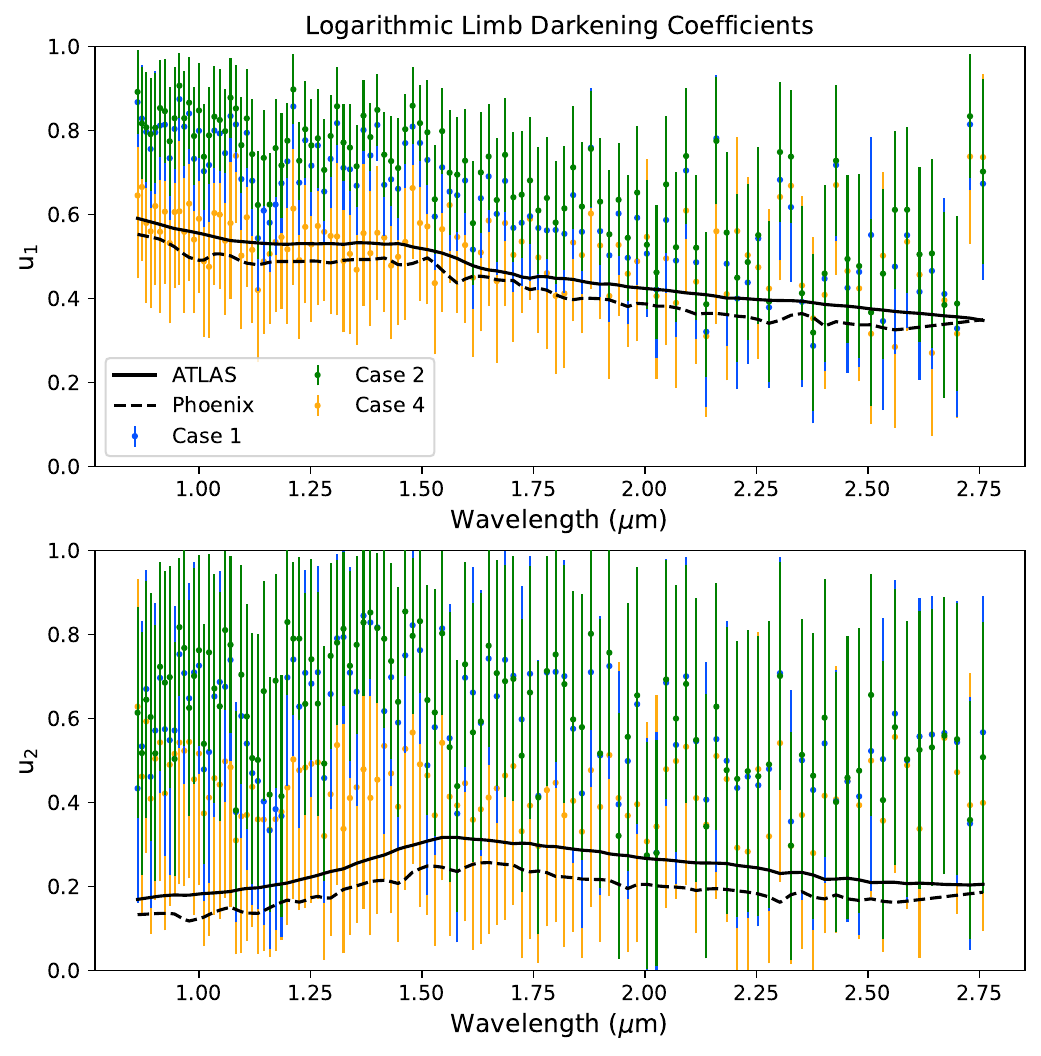}
    \caption{Comparing the best-fit logarithmic limb darkening coefficients $u_1$ (top) and $u_2$ (bottom) from Cases 1, 2, and 4 to predictions from ATLAS (solid) and Phoenix (dashed) stellar models.
    \href{https://github.com/kronos-jwst/KRONOS-I-V1298-Tau-c-NIRISS/blob/main/Create_Figure12.ipynb}{\githubicon}}
    \label{fig:LDCcomp}
\end{figure*}

We also compared our freely fitted limb darkening coefficients to predictions from stellar models. Figure~\ref{fig:LDCcomp} shows the best-fit logarithmic $u_1$ (top panel) and $u_2$ (bottom panel) values from Cases 1 (blue), 2 (green), and 4 (orange). We compare to ATLAS (black solid) and Phoenix (black dashed) model predictions, each computed via ExoCTK. 
While additional stellar models, such as the Stagger grid \citep{Stagger_grid}, are commonly used in the exoplanet literature as well, logarithmic coefficients from these are not readily available from community tools.
In all cases, the freely fit coefficient spectra follow the same shape as the model predictions, suggesting we indeed are fitting the limb darkening profile instead of light curve noise, but the values are vertically offset. 
Relative to the ATLAS models, the median differences for $u_1$ were 0.17, 0.23, and 0.03 for Cases 1, 2, and 4, respectively. For $u_2$, these were 0.35, 0.39, and 0.19. The differences are largely wavelength independent, particularly for $u_1$. On the other hand for $u_2$, there are slightly exaggerated differences around 1.25 - 1.5~$\mu$m but not more than $\sim$1$\sigma$ per channel. This appears to reflect a difference in where the coefficient profile turns over, as the models predict the peak value to occur around $\sim$1.55~$\mu$m but we instead measure it near $\sim$1.4~$\mu$m.
Evident by the error bar sizes in each panel, we also find that the $u_1$ coefficient is much better constrained by the data than $u_2$.

\begin{figure*}
    \centering
    \includegraphics[width=\linewidth]{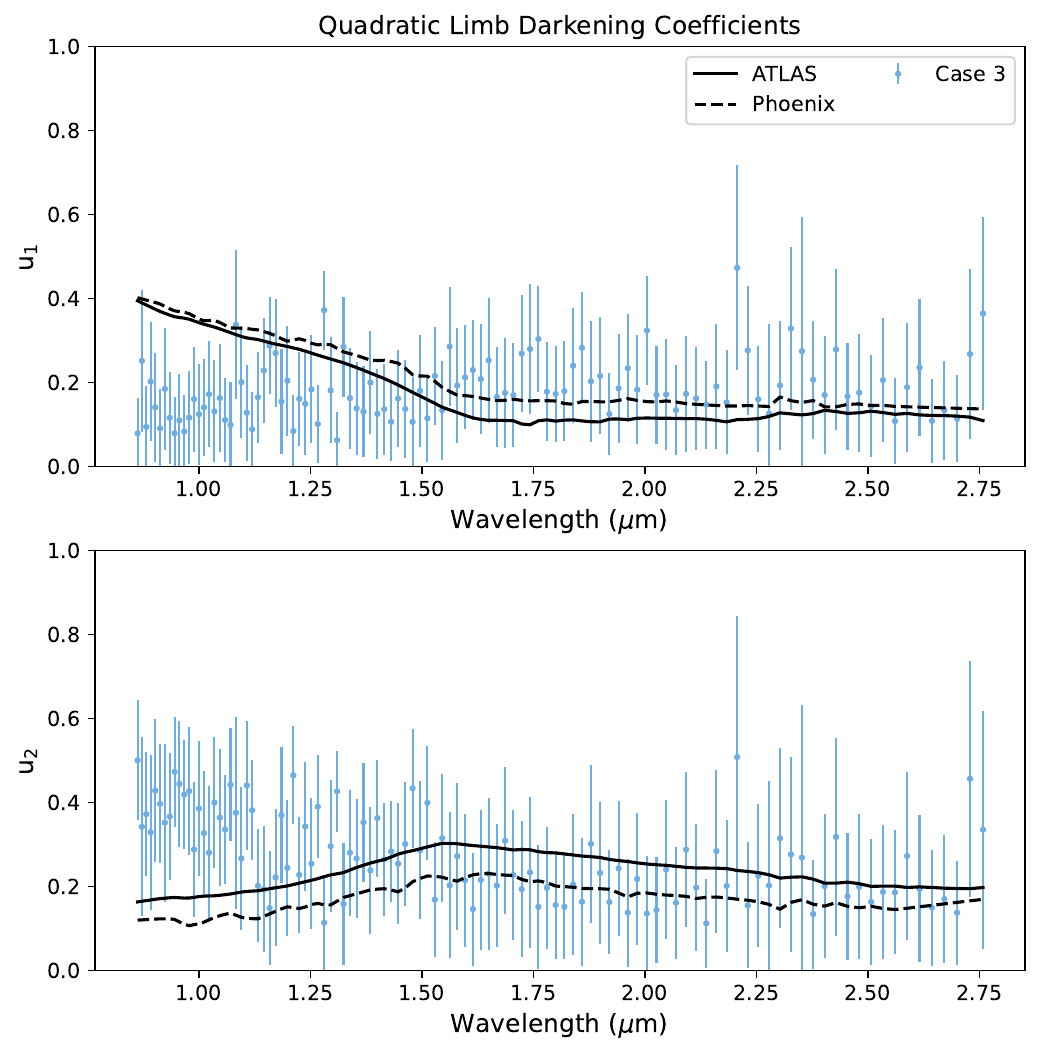}
    \caption{Comparing the best-fit quadratic limb darkening coefficients $u_1$ (top) and $u_2$ (bottom) from Cases 3 to predictions from ATLAS (solid) and Phoenix (dashed) stellar models.
    \href{https://github.com/kronos-jwst/KRONOS-I-V1298-Tau-c-NIRISS/blob/main/Create_Figure13.ipynb}{\githubicon}}
    \label{fig:LDCcomp_quad}
\end{figure*}

Figure~\ref{fig:LDCcomp_quad} presents the same comparison but for quadratic coefficients from Case 3. We see less consistency between the freely fitted values and the model predictions, particularly in the shape of the coefficient spectrum. Both models predict an upward slope in $u_1$ below $\sim$1.55~$\mu$m, but the freely fitted coefficients are instead constant in this regime. Similarly, the model values in this regime slope downward in $u_2$, but the freely fitted values instead have a mild upward slope. At longer wavelengths where the profiles all flatten, the freely fitted coefficients are fairly consistent in magnitude with the Phoenix model predictions, and slightly offset relative to the ATLAS values. 

\cite{kostogryz2024_limbdarkening} recently showed that a star's magnetic field can significantly alter its limb darkening profile, and argued this is likely a key reason for previously observed discrepancies between measured and predicted values \citep[e.g.,][]{maxted2018_limbdarkening, maxted2023_limbdarkening, rustamkulov2023_wasp39bERS, sing2024_wasp107, schlawin2024_gj1214}. The limb darkening coefficients derived from stellar models, such as those shown above in Figures~\ref{fig:LDCcomp} and \ref{fig:LDCcomp_quad}, typically assume the star has no magnetic field \citep{kostogryz2024_limbdarkening}. Previous observations of V1298~Tau suggest that V1298~Tau not only has a significant magnetic field strength, but that its strength may vary by nearly $\sim$2$\times$ over a few years \citep{finociety_2023}. Therefore, the magnetic field of V1298~Tau is likely driving the observed discrepancies in the limb darkening coefficients. 

\section{Full Atmospheric Retrieval Results} \label{apx:retrievalresults}

\begin{figure*}
    \centering
    \includegraphics[width=\linewidth]{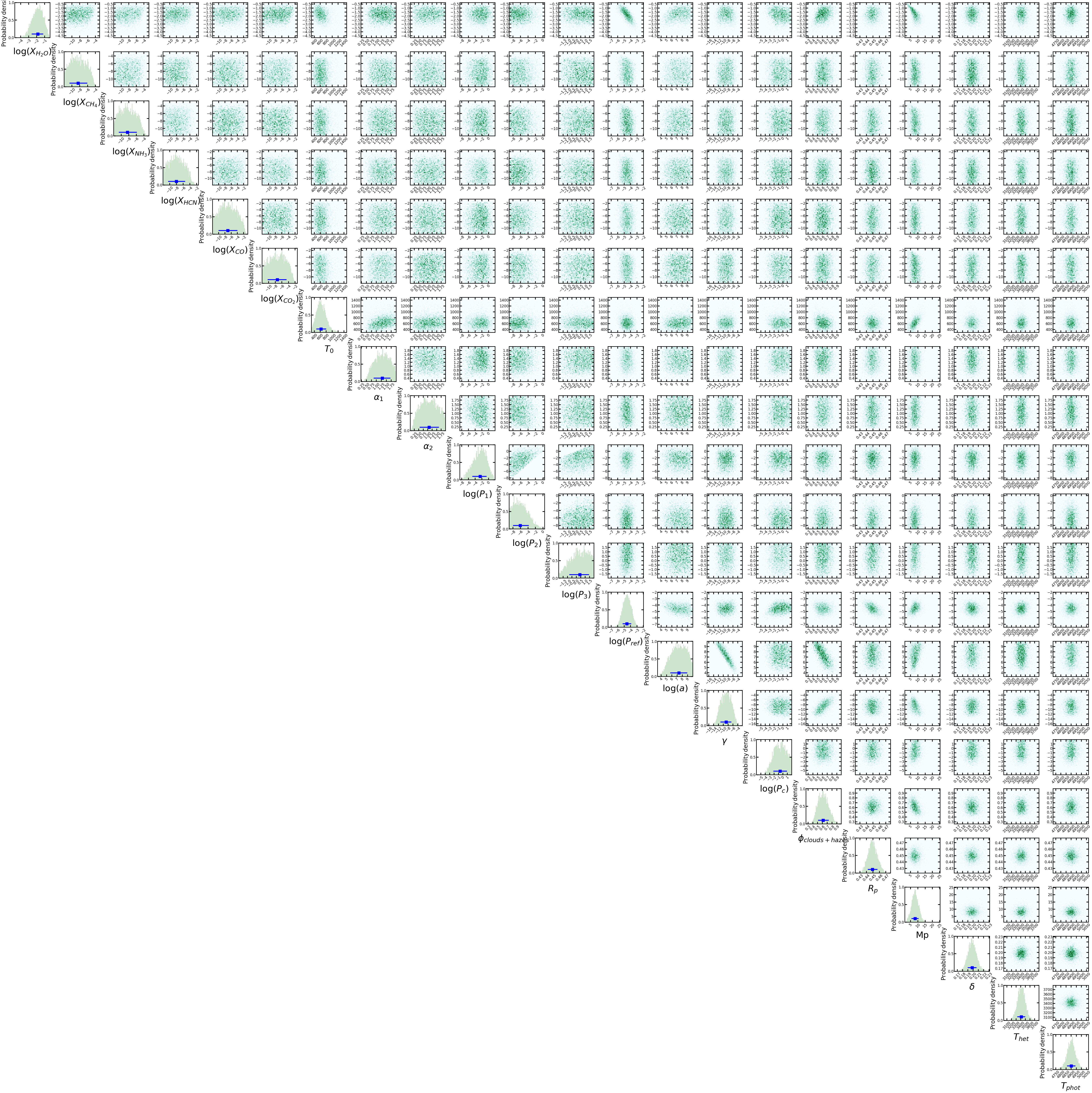}
    \caption{Corner plot from our retrieval using informed stellar heterogeneity priors.}
    \label{fig:cornerplot_informedpriors}
\end{figure*}

\begin{figure*}
    \centering
    \includegraphics[width=\linewidth]{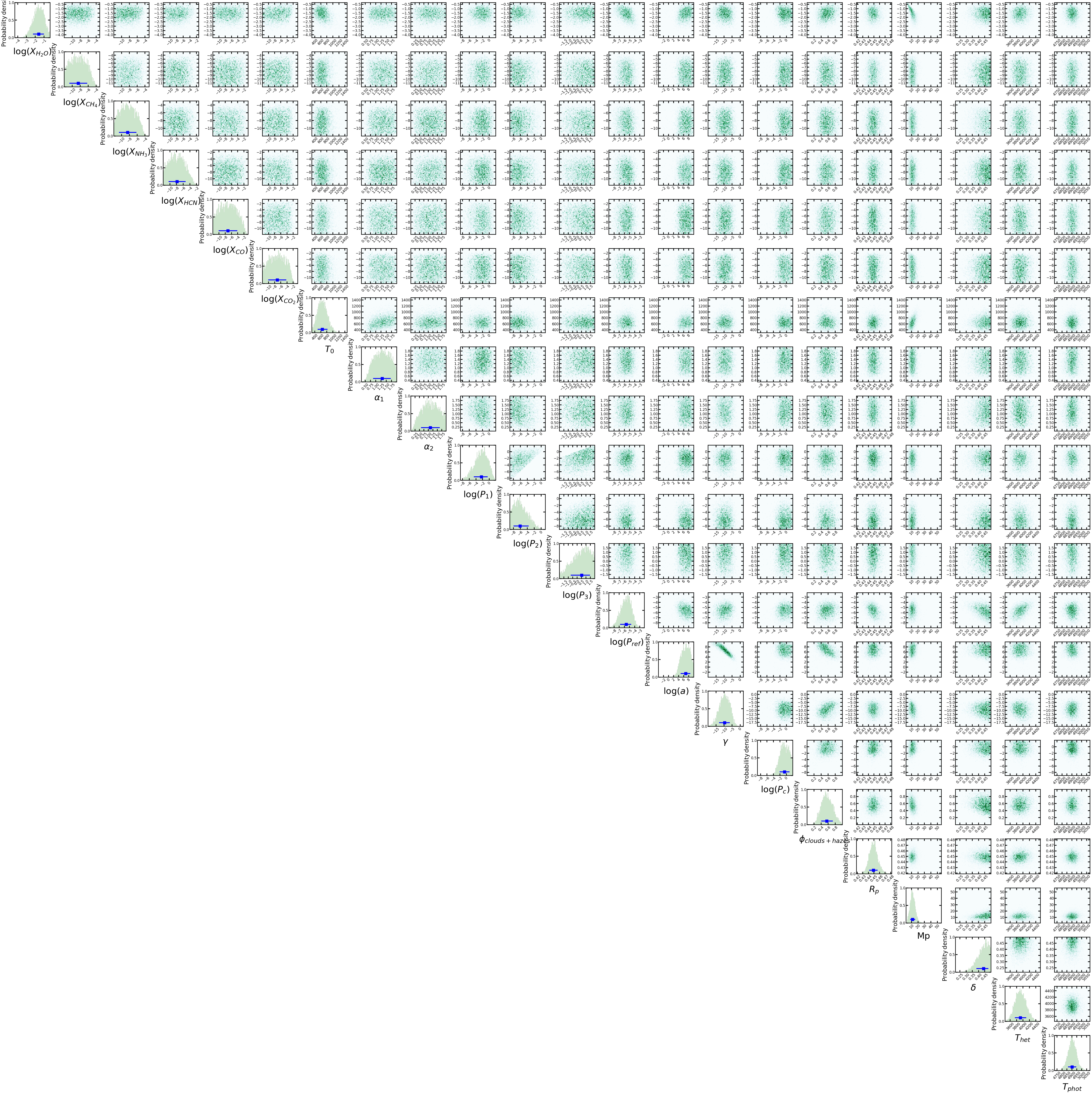}
    \caption{Corner plot from our retrieval using uninformed stellar heterogeneity priors.}
    \label{fig:cornerplot_uninformedpriors}
\end{figure*}

Here we provide the full posterior corner plots from our atmospheric retrievals with informed priors (Figure~\ref{fig:cornerplot_informedpriors}) and with uninformed priors (Figure~\ref{fig:cornerplot_uninformedpriors}) on the stellar heterogeneity parameters.

\section{Collection of Literature Metallicities} \label{apx:metallicities}

In Section~\ref{subsec:formation_massmetallicity}, we present a collection of metallicities for mature exoplanets. These are based on a variety of methodologies, and are inferred from the presence and estimated abundances of various combinations of molecules: 
H$_2$O  \citep[GJ~9827~d and TOI-421~b;][]{piaulet2024_GJ9827d, davenport2025}, 
CH$_4$ \citep[LP791-18~c;][]{roy2026_lp79118},
SO$_2$ + H$_2$S \citep[L98-59~d;][]{gressier2024_l9859d},
H$_2$O + CO$_2$ \citep[WASP-166~b;][]{mayo2025_wasp166b}, 
H$_2$O + CO$_2$ + CO \citep[WASP-69~b;][]{schlawin2024_wasp69b},
H$_2$O + CO$_2$ + CH$_4$ \citep[TOI-270~d;][]{benneke2024_toi270d},
H$_2$O + CO$_2$ + SO$_2$ \citep[HAT-P-26~b;][]{gressier2025_hatp26b},
H$_2$O + CO$_2$ + CH$_4$ + SO$_2$ \citep[GJ~3470~b;][]{beatty2024_gj3470},
H$_2$O + CO$_2$ + CH$_4$ + SO$_2$ + CO + NH$_3$ \citep[WASP-107~b;][]{Welbanks2024}. We also include K2-18~b whose exact atmospheric composition is currently the subject of intense debate. We use the metallicity estimate of \cite{jaziri2025_k218b} which is broadly consistent with the results of numerous independent analyses \citep{mahdu2023, wogan2024, Welbanks2025, schmidt2025_k218b, lavvas2026_k218b}.

The lower limits are derived by \cite{Moran2023_GJ486b, schlawin2024_gj1214, alderson2024_toi836b, cadieux2024_lhs1140, scarsdale2024_L9859c,  ahrer2025_gj3090, alam2025_l1689, belloarufe2025_l9859b, kahle2025_hd86226c, luque2025_TOI1685b, meech2026_toi260b, redai2025_gj357b, rigby2025_toi732c, teske2025_TOI561b,  teske2025_toi776c, wallack2026_hd15337c}.

%%%%%%%%%%%%%%%%%%%%%%%%%%%%%%%%%%%%%%%%%%%%%%%%%%

\bibliography{refs}{}
\bibliographystyle{aasjournalv7}



\end{document}